\documentclass[nofootinbib,preprint,
  showpacs,preprintnumbers,amsmath,amssymb]{revtex4}

\usepackage{latexsym}

\newcommand{\fal}[2]{\frac{\delta{#1}}{\delta{#2}}} % abbreviation for functional derivatives
\newcommand{\zfal}[3]{\frac{\delta^2{#1}}{\delta{#2}\delta{#3}}}  % second functional derivative
\newcommand{\threenabla}{\,^{3}\nabla}              % 3d covariant derivative with torsion
\newcommand{\threesD}{{\,}^{\rm{3s}}\hspace{-0.0mm}\mathcal{D}} % 3d spinorial cov. derivative without torsion
\newcommand{\threeD}{{\,}^{3}\hspace{-0.0mm}\mathcal{D}}  % 3d spinorial cov. derivative with torsion
\newcommand{\threeR}{{\,}^{3}\hspace{-0.5mm}R}      % 3d curvature with torsion
\newcommand{\threesR}{{\,}^{\rm{3s}}\hspace{-0.5mm}R} %% 3d curvature without torsion
\newcommand{\Hbot}{\mathcal{H}_{\bot}}
\newcommand{\Hmbot}{{\mathcal{H}^{\rm{m}}_\bot}}
\newcommand{\eps}{\epsilon}
\newcommand{\DD}{D_{B\hspace{1.5mm}mj}^{\hspace{1.5mm}B'}D^C_{\hspace{1.5mm}A'kl}}
\newcommand{\threeom}{\,^3\hspace{-0.2mm}\omega} %3d spin connection with torsion
\newcommand{\threesom}{\,^{3\rm{s}}\hspace{-0.2mm}\omega} %3d spin conn. without t.
\newcommand{\threebarom}{\,^3\hspace{-0.2mm}\bar{\omega}} %conjugate 3d spin conn. with t.
\newcommand{\threekappa}{\,^{3}\hspace{-0.2mm}\kappa} %3d contorsion
\newcommand{\threesbarR}{{\,}^{\rm{3s}}\hspace{-0.5mm}\bar{R}}  %conjugate comp. of 3d curvature without torsion
\newcommand{\threebarR}{{\,}^{3}\hspace{-0.5mm}\bar{R}} %conjugate comp. of 3d curvature with torsion
\newcommand{\threesbarom}{\,^{3\rm{s}}\hspace{-0.2mm}\bar{\omega}} %conjugate comp. of 3d spin conn.
\newcommand{\bpsi}{\bar{\psi}}                      % conjugate gravitino
\newcommand{\E}{E^{CAB'i}_{Bjk}}
\newcommand{\DBij}{D^{BB'}_{ij}}
\newcommand{\DBAli}{D^{B}_{\hspace{1.5mm}A'li}}
\newcommand{\DBAlk}{D^{B}_{\hspace{1.5mm}A'lk}}

\begin{document}

\title{\bf Semiclassical approximation to supersymmetric
 quantum gravity}
\author{Claus Kiefer}
\email{kiefer@thp.uni-koeln.de} \affiliation{ Institut f\"ur Theoretische Physik, Universit\"at zu K\"oln,
Z\"ulpicher Str. 77, 50937 K\"oln, Germany.}
\author{Tobias L\"uck}
\email{lueck@thp.uni-koeln.de} \affiliation{ Institut f\"ur Theoretische Physik, Universit\"at zu K\"oln, Z\"ulpicher
Str. 77, 50937 K\"oln, Germany.}
\author{Paulo  Moniz}\thanks{On leave of absence from
Departmento de Fisica, UBI,
  Covilha, Portugal;
Also at Centra--IST,  Lisboa,
  Portugal;  URL: {\tt http://webx.ubi.pt/$\sim$pmoniz}}~
\email{p.moniz@qmul.ac.uk}
\affiliation{Astronomy Unit, School of Mathematical Sciences,
Queen Mary College, University of London,
Mile End Road, London E1 4NS, United Kingdom.}
%\date{}
\begin{abstract}
We develop a semiclassical approximation scheme for the
constraint equations of supersymmetric canonical quantum gravity.
This is achieved by a Born--Oppenheimer type of expansion,
in analogy to the case of the usual Wheeler--DeWitt equation.
The formalism is only consistent if the states at each order
depend on the gravitino field. We recover at consecutive orders
the Hamilton--Jacobi equation, the functional Schr\"odinger equation,
and quantum gravitational correction terms to this Schr\"odinger equation.
In particular,
the following consequences are found:
 $(i)$ the Hamilton--Jacobi equation and therefore the
background spacetime must involve the gravitino,
$(ii)$ a (many fingered) local time parameter has to be
present on $SuperRiem~\Sigma$ (the space of
all possible tetrad and gravitino fields), $(iii)$
 quantum supersymmetric gravitational corrections
affect the evolution of the very early universe.
The physical meaning of these equations and results,  in particular
the similarities to and differences from the pure bosonic case, are
discussed.
\end{abstract}
 %% Semiclassical theories in gauge fields ???

\maketitle

%\input{Intro.tex}
%\input{Quantization.tex}
%\input{Scheme.tex}
%\input{HamiltonJacobi.tex}
%\input{Schroedinger.tex}
%\input{Corrections.tex}
%\input{Conclusions.tex}
%\input{Appendix.tex}

%%%%%%%%%%%%%%%%%%%%%%%%%%%%%%%%%%%%%%%%%%%%%%%%%%%%%%%%%%%%%%%%%%%%%%%%%%%%%%%%%%%%%%%%%%%%%%%%

\section{Introduction\label{intro}}

The consistent accomodation of the gravitational interaction into the framework of a quantum theory remains
to be completed. Among the major approaches are string theory and canonical quantum gravity, cf.
\cite{OUP,CQg,SSt} and the references therein. It is generally assumed that supersymmetry \cite{SUSY} is a
major ingredient of string theory. This is also one of the motivations for the study of a supersymmetric
version of the canonical quantization of gravity, independent of the search for a unified theory of all
interactions. In addition, the implementation of supersymmetry (SUSY) into canonical quantum gravity may
simplify the formalism in a specific aspect: If the quantum constraint algebra closes, the (more complicated)
Hamiltonian constraint is automatically fulfilled once the (simpler) SUSY constraints hold. The formalism is
especially simplified if suitable choices are made regarding the canonical conjugate momenta, leading to
simplifications in the SUSY constraints and therefore the Hamiltonian constraint. This has led to a detailed
study of supersymmetric canonical quantum gravity, see \cite{Death,Paulo} and the references therein for an
introduction and review. Pertinent applications include 
black-hole physics \cite{Moniz:1996qz} and quantum
cosmology \cite{Death,Paulo,SQC1}.

Research in supersymmetric quantum cosmology (SQC)
provides the means to investigate some relevant problems concerning the 
evolution of the very early universe, namely ($a$) relating 
exact solutions found for spatially isotropic and anisotropic 
cosmologies with those obtained from the use of 
specific boundary conditions in usual quantum cosmology
\cite{SQC2}, ($b$) probing how 
the symmetry properties of dualities
in  superstring theory can be induced into 
the quantum states \cite{SQC3} and 
($c$) analyzing the possibility of inflation occurrence and 
structure formation \cite{SQC4a,SQC4b,Z14}.
The essential feature is that 
SQC  subscribes to the idea that treating both
quantum gravity
and SUSY effects as dominant would give an improved
description of the early universe.
Such an approach can bring 
profound consequences for the wave function
of the universe: the quantum state  can be 
written as an expansion in linearly independent fermionic sectors, 
each  associated with a specific bosonic functional 
(of the same type as those satisfying the 
Wheeler--DeWitt equation).  Besides the 
pertinent question of how to 
interpret the meaning of such quantum states 
\cite{Swfu1},  
investigating   whether conserved currents and 
a  positive 
probability density can be obtained 
in this setting \cite{Swfu2}
must be performed by taking into account the 
enlarged structure for the wave function. 
 Moreover, any
 cosmological evolution  
 determined within SQC 
 must eventually 
be consistent with a mechanism 
for SUSY breaking  
\cite{SQC4a,Swfu3}.
Nevertheless, in spite of all the 
progress achieved so far, 
 further efforts are required to  find {\em new}  states
 determining a consistent dynamical path 
 from a supersymmetric quantum cosmological 
to a  classical cosmological 
stage \cite{Paulo}.

It is in the context of the above description that the
purpose of the present paper can be seen: investigate
the semiclassical approximation
of supersymmetric canonical quantum gravity. The viability of such
a scheme is crucial for the framework of
quantum gravity in the
presence of supersymmetry. In the bosonic case,
a formal Born--Oppenheimer type of approximation scheme
has been successfully applied to the Wheeler--DeWitt equation \cite{OUP}.
 This has led in particular to the derivation of
quantum gravitational corrections terms which modify the limit
of quantum theory on a fixed background spacetime \cite{KS91,Honnef,BK}.
Here we extend this formalism to the supersymmetric case.
We restrict ourselves to the theory of $N=1$ Supergravity
(SUGRA) in four spacetime dimensions.\footnote{$N=1$ SUGRA
in four spacetime dimensions is the simplest
SUSY extension of general relativity.
It is related to the theory
of $N=1$ SUGRA in eleven spacetime dimensions, to which
superstrings are associated in the
context of M-theory.}
Compared with the case of the bosonic Wheeler--DeWitt equation,
 this leads to equations of the same type, bearing similarities
 as well as various
important differences arising from the
presence of fermions (via SUSY).
The framework and results presented herewith
may constitute an efficient means to
study the influence of SUSY in the physics of the
very early universe.

Our article is hence organized as follows.
 In Section~II the Hamilton--Jacobi
equation for supergravity is recovered. It is shown that the presence of the
gravitino field is mandatory at each order of the approximation.
The gravitino thus has to appear already at the order of the classical
background spacetime. Section~III presents the derivation of the
functional Schr\"odinger equation for non-gravitational fields, that is,
the limit of quantum field theory on a given background. In Section~IV
we then derive supersymmetric quantum gravitational
correction terms to this equation.
Section~V conveys a  summary and discussion of our work and
results, together
with an outlook of subsequent 
future reasearch to be 
followed. In order to assist the
reader, some Appendices have been included.
In Appendix~A we review the
formalism of supersymmetric canonical quantum gravity. The Born--Oppenheimer
scheme for the bosonic case is briefly reviewed in Appendix~B.
Technical calculations
referring to
Sections~II--IV are relegated to Appendices C and D.

%%%%%%%%%%%%%%%%%%%%%%%%%%%%%%%%%%%%%%%%%%%%%%%%%%%%%%%%%%%%%%%%%%%%%%%

\section{Recovery of the Hamilton--Jacobi equation\label{HamiltonJacobi}}

The purpose of this section is the application of the
semiclassical approximation scheme, previously developed for
the non-supersymmetric case, to canonical quantum gravity with SUSY.
Readers who are not familiar with semiclassical gravity, or with
the canonical formalism for SUSY quantum gravity,
may wish to consult Appendices A, B, and C, see also \cite{Lueck}.

We start by mentioning a well known 
general feature of all supersymmetric theories,
namely  that the commutator of a primed and an unprimed SUSY
transformation yields a coordinate transformation in spacetime.
This translates into the anticommutator expression 
\begin{equation}
\label{Ncommutator of S_A, S_A'}
  [S_A(x),\bar{S}_{A'}(y)]_+ =4\pi G\hbar\mathcal{H}_{AA'}(x)\delta(x,y)\ ,
\end{equation}
where $S_A(x)$ and $\bar{S}_{A'}(y)$ are the constraints 
corresponding to the SUSY transformations. 
For consistency, then, $\mathcal{H}_{AA'}$ should also vanish 
as a constraint. This constraint is in fact related to the generators
of spacetime transformations. 
We shall make use of the decomposition
\begin{equation}
\label{H_AA, quantum, decomposition}
  \mathcal{H}_{AA'}=-n_{AA'}\mathcal{H}_{\bot}+e_{AA'}^i\mathcal{H}_i\ ,
\end{equation}
where $n_{AA'}$ and $e_{AA'}^i$ are the spinorial versions of
the normal vector and the dreibein, respectively, cf.
Appendix~A;
$\mathcal{H}_i$ and $\mathcal{H}_{\bot}$ denote, 
respectively, the gravitational momentum 
and Hamiltonian constraints. 
In particular, $\Hbot$ is the normal
projection of the constraint $\mathcal{H}_{AA'}$.
We obtain it from (\ref{H_AA, quantum, decomposition})
after multiplication and contraction with $-n^{{AA'}}$.
Explicitly we have (in a quantum mechanical representation; 
the quantities are introduced and explained in
Appendix~A),
\begin{eqnarray}
  \label{constraint, quantized HAA}
    \mathcal{H}_{AA'} &=&  4\pi G i
    \hbar^2\psi^B_i\fal{}{e^{AB'}_j}\left[\eps^{ilm}D_{B\hspace{1.5mm}mj}^{\hspace{1.5mm}B'}D^C_{\hspace{1.5mm}A'kl}
    \fal{}{\psi^C_k}\right]
  \nonumber\\ &&
    -4\pi G i\hbar^2\fal{}{e^{AB'}_j}\left[\DBij\fal{}{e^{BA'}_i}\right]
  \nonumber\\ &&
    - \frac{i\hbar}{2}\eps^{ijk}\left[\left({\threesD}_j\psi_{Ak}\right)\DBAli\fal{}{\psi^B_l}+
    \psi_{Ai}\left({\threesD}_j\DBAlk\fal{}{\psi^B_l}\right)\right]
  \nonumber\\ &&
    - {i\hbar\,\threesD}_i\left(\fal{}{e^{AA'}_i}
    +\frac{1}{2}\eps^{ijk}\psi_{Aj}D^{B}_{\hspace{1.5mm}A'lk}\fal{}{\psi^B_l}\right)
    +n_{AA'}\frac{1}{G}V[e]\ ,
\end{eqnarray}
where $V[e]=\sqrt{h}\threesR/16\pi$.
For the semiclassical
approximation  developed here we employ  this  version
(\ref{Ncommutator of S_A, S_A'})--(\ref{constraint, quantized HAA})
of the constraints instead of those
extracted directly from  the action, cf. \cite{Death,Paulo}.
This has the
advantage that the (formal) closure of the algebra, cf. also Ref.
(\ref{commutator of S_A, S_A'}), is automatically implemented.
Expression 
(\ref{constraint, quantized HAA})
can be written in a less symmetric, but somewhat simplified
form.
For the fermionic part of the last line in (\ref{constraint, quantized HAA}),
one can write
\begin{eqnarray}
    -\frac{1}{2}{i\hbar\,\threesD}_i\left(\eps^{ijk}\psi_{Aj}D^{B}_{\hspace{1.5mm}A'lk}\fal{}{\psi^B_l}\right)
    &=&\frac{1}{2}\eps^{ijk}\threesD_i(\psi_{Aj}\bar{\psi}_{A'k})
  \nonumber\\
    &=&\frac{1}{2}\eps^{ijk}\left[(\threesD_i\psi_{Aj})\bar{\psi}_{A'k}
    +\psi_{Aj}(\threesD_i\bar{\psi}_{A'k})\right]
  \nonumber\\
    &=&\frac{1}{2}\eps^{ijk}\left[(\threesD_j\psi_{Ak})\bar{\psi}_{A'i}
    -\psi_{Ai}(\threesD_j\bar{\psi}_{A'k})\right]\,.
  \nonumber
\end{eqnarray}
Comparing this with the third line of (\ref{constraint, quantized HAA}),
\begin{eqnarray*}
  &-& \frac{i\hbar}{2}\eps^{ijk}\left[\left({\threesD}_j\psi_{Ak}\right)D^{B}_{\hspace{1.5mm}A'li}\fal{}{\psi^B}_{l}+
    \psi_{Ai}\left({\threesD}_jD^{B}_{\hspace{1.5mm}A'lk}\fal{}{\psi^B_l}\right)\right]
  \nonumber\\
  &&\hspace{15mm}=\frac{1}{2}\eps^{ijk}\left[(\threesD_j\psi_{Ak})\bar{\psi}_{A'i}
    +\psi_{Ai}(\threesD_j\bar{\psi}_{A'k})\right]\,,
  \nonumber
\end{eqnarray*}
we find that the terms containing $\threesD_j(\bar{\psi}_{A'k})$ cancel out. The normal projection of the
remaining term containing $\threesD_j\psi_k^A$ is given by,
cf. (\ref{A15}) and (\ref{Quantum repr. of the momenta and fields}),
\begin{eqnarray*}
    n^{AA'}{i\hbar}\eps^{ijk}\left({\threesD}_j\psi_{Ak}\right)D^{B}_{\hspace{1.5mm}A'li}
    \fal{}{\psi^B_{l}}&=&
    \frac{2\hbar}{\sqrt{h}}n^{AA'}e^{BC'}_ie_{CC'l}n^{C}_{\hspace{1.6mm} A'}
    \eps^{ijk}\left({\threesD}_j\psi_{Ak}\right)\fal{}{\psi^B_l}
  \nonumber\\
    &=&\frac{\hbar}{\sqrt{h}}\eps^{ijk}e^{BC'}_ie^{A}_{\hspace{1.6mm}
    C'l}\left({\threesD}_j\psi_{Ak}\right)\fal{}{\psi^B_l}\,.
  \nonumber
\end{eqnarray*}
For later use we introduce
for the normal projection of the expression $\eps^{ilm}\DD$
the definition,
\begin{eqnarray}
\label{E, definition}
    E^{CAB'i}_{Bjk}\equiv n^{AA'}\eps^{ilm}\DD&=&
    \frac{-2i}{\sqrt{h}}\eps^{ilm}D_{B\hspace{1.6mm} mj}^{\hspace{1.6mm} B'}e^{CD'}_le_{DD'k}n^D_{A'}n^{AA'}
  \nonumber\\
    &=&\frac{i}{\sqrt{h}}\eps^{ilm}D_{B\hspace{1.6mm} mj}^{\hspace{1.6mm} B'}e^{CD'}_le^{A}_{D'k}\ .
\end{eqnarray}
We then get the following expression from (\ref{constraint, quantized HAA}),
\begin{eqnarray}
\label{Hbot}
    \mathcal{H}_{\bot}  =  -n^{{AA'}}\mathcal{H}_{{AA'}}  &= &
    \underbrace{-4\pi i G \hbar^2 n^{AA'}\psi^B_i\fal{}{e^{AB'}_j}\left[\eps^{ilm}\DD\fal{}{\psi^C_k}\right]}_{\rm{(i)}}
  \nonumber\\&&
    \underbrace{+4\pi G i\hbar^2n^{AA'}\fal{}{e^{AB'}_j}\left[D^{BB'}_{ij}\fal{}{e^{BA'}_i}\right]}_{\rm{(ii)}}
  \nonumber\\&&
    \underbrace{+\frac{\hbar}{\sqrt{h}}\eps^{ijk}e^{BC'}_ie^{A}_{\hspace{1.6mm} C'l}
    \left({\threesD}_j\psi_{Ak}\right)\fal{}{\psi^B_l}}_{\rm{(iii)}}
  \nonumber\\&&
    \underbrace{+{i\hbar n^{AA'}\,\threesD}_i\fal{}{e^{AA'}_i}}_{\rm{(iv)}}
    \underbrace{-\frac{1}{G}V[e]}_{\rm{(v)}}\ .
\end{eqnarray}
Since in the parts (i) and (ii) of (\ref{Hbot}) the functional derivative $\delta/\delta e^{AB'}_j$ can also act
on $\DD$ and $D^{BB'}_{ij}$, we first calculate these derivatives (see Appendix \ref{APP: Formulas}). We find
\begin{eqnarray}
\label{derivative of epsDD}
    n^{AA'}\eps^{ilm}\fal{}{e^{AB'}_j}\DD=
   \left(\frac{-3i}{\sqrt{h}}\delta^i_k
\eps_B^{\hspace{1.6mm} C}-2h^{ij}\eps_{jkl}n^{CB'}e_{BB'}^l\right)\delta(0)\ ,
\end{eqnarray}
and
\begin{equation}
\label{derivative of DBij}
  n^{{AA'}}\fal{}{e^{AB'}_j}D^{BB'}_{ij}
  =\frac{i}{\sqrt{h}}e^{BA'}_i\delta(0)\ .
\end{equation}
The `divergence' $\delta(0)$ arises from the functional derivative at
the same space point. It has to be regularized in a rigorous way, which
is beyond the scope of this article. For the semiclassical approximation,
addressing this issue is of less relevance, since it just corresponds to a
factor ordering ambiguity. In the rest of this article we shall suppress
$\delta(0)$.

The SUSY version of
the Wheeler--DeWitt equation is then found to read
\begin{equation}
\label{Hamiltonian constraint, formal}
  (\Hbot+\Hmbot)\Psi=0\ ,
\end{equation}
where $\Hbot$ (see discussion above) denotes
the gravitational SUSY contribution to the Hamiltonian constraint, and
$\Hmbot$ is the contribution from non-gravitational (`matter') fields.
For definiteness we shall take for
$\Hmbot$ the Hamiltonian density of a
minimally coupled scalar field $\Phi$,
\begin{equation}
  \label{matter Hamiltonian, explicit}
  \Hmbot = \frac{1}{2}\left(-\frac{\hbar^2}{\sqrt{h}}\fal{^2}{\Phi^2}
  +\sqrt{h}h^{ij}\Phi_{,i}\Phi_{,j}+\sqrt{h}(m^2\Phi^2+U(\Phi))\right)\ ,
\end{equation}
where $h_{ij}$ denotes the three-metric, $h$ its determinant,
and the self-coupling potential $U(\Phi)$ is left unspecified.
The more realistic (and more complicated) case
of supermatter should be treated in a future work, cf.
\cite{Death,Paulo}.
The   state $\Psi$ of SUSY quantum gravity is a wave functional defined
on the space of all tetrad and gravitino fields
(plus possible other fields) on a spatial
hypersurface $\Sigma$. We shall
call this space $SuperRiem~\Sigma$,
extending the notion $Riem~\Sigma$ for the space of all three-metrics in
canonical quantum gravity \cite{OUP}.

Similarly to  the bosonic case, we shall use an  ansatz
of the form (see Appendix B)
\begin{equation}
  \Psi[e,\psi,\Phi]=\exp\left(\frac{i}{\hbar}S[e,\psi,\Phi]\right)\ ,
  \label{NEW20}
\end{equation}
and expand $S$ into a power series with respect to $G$,
\begin{equation}
  S[e,\psi,\Phi]=\sum_{n=0}^{\infty}S_n[e,\psi,\Phi] G^{n-1}\ .
  \label{NEW21}
\end{equation}
By means of this procedure, we then
investigate  the expansion of (\ref{Hamiltonian constraint, formal}) in powers of $G$.
The lowest order is
$G^{-2}$. As in the bosonic case, this yields the independence of
$S_0$ on the matter field $\Phi$, that is, $S_0\equiv S_0[e,\psi]$,
cf. Appendix~B.

At order $G^{-1}$ we find contributions which determine the Hamilton--Jacobi equation of supersymmetric quantum gravity. It reads
\begin{eqnarray}
  \label{susy HJ eq.}
  0 &=&  4\pi i\left(\psi^B_i\fal{S_0}{e^{AB'}_j}{E^{CAB'i}_{Bjk}}\fal{S_0}{\psi^C_k}
  -n^{AA'}{D^{{BB'}}_{ij}}\fal{S_0}{e^{AB'}_j}\fal{S_0}{e^{BA'}_i}\right)
  \nonumber\\
  &&+\frac{i}{\sqrt{h}}\eps^{ijk}e^{BC'}_ie^{A}_{\hspace{1.6mm} C'l}
    \left({\threesD}_j\psi_{Ak}\right)\fal{S_0}{\psi^B_l}
    -n^{AA'}\threesD_i\fal{S_0}{e^{AA'}_i}-V\,.
\end{eqnarray}
Let us begin by
investigating the question whether the Hamilton--Jacobi
equation,
\begin{equation}
  \label{bos. HJ eqN.}
  \frac{1}{2}G_{ijkl}\fal{S_0}{h_{ij}}\fal{S_0}{h_{kl}}+V^{\rm g}=0\ ,
\end{equation}
of the bosonic case
is contained in (\ref{susy HJ eq.}). For this purpose we ignore all terms involving the
gravitino and reformulate the remaining part in terms of the three-metric $h_{ij}$.

If we assume that $S_0[e]$ can be rewritten as $S_0[h_{ij}]$, we can use
Equation (\ref{APP: trafo
from e to h}) to transform the
functional derivatives. In addition, we have to take into account that the
expansion parameter used in Appendix \ref{Scheme}
(where the conventions of the bosonic case are used)
differs by a factor $32\pi$
 from the one we use here. Moreover, we have the following relations,
\begin{equation}
  V[e]=-\frac{1}{32\pi}V^{\rm{bos}}[h_{ij}]\,,
\end{equation}
and
\begin{equation}
  S_0=\frac{1}{32\pi}S_0^{\rm{bos}}\,.
\end{equation}
This, then, leads to
\begin{equation}
\label{HJ}
  \frac{1}{2}G_{ijkl}\fal{S_0^{\rm{bos}}}{h_{ij}}\fal{S_0^{\rm{bos}}}{h_{kl}}
  +64\pi n^{AA'}\partial_i e_{AA'j}\fal{S_0^{\rm{bos}}}
{h_{ij}}+V^{\rm{bos}}=0\,.
\end{equation}
A comparison with (\ref{bos. HJ eqN.}) shows that there is almost
a total equivalence. Only the second
term in (\ref{HJ}) has no counterpart there.
Its presence can be traced back to the
following observation.

As described in detail in \cite{Death,Paulo}, the 
constraints of supersymmetric quantum gravity can be
established either $(i)$ directly from the $N=1$ SUGRA action 
by means of a variational principle 
or $(ii)$ by further 
simplifying those 
mathematical expressions 
through a sensible choice of, for example,  the canonical conjugate momenta in
their (quantum) operator representation; this leads to 
much more tractable expressions. 
It is the anticommutator of 
the SUSY constraints within ($ii$)
that produces the much simpler form of
 $\mathcal{H}_{AA'}$  
and therefore of
$\mathcal{H}_{\bot}$ and $\mathcal{H}_i$ 
 (see Eqs. 
  (\ref{Ncommutator of S_A, S_A'})--(\ref{constraint, quantized HAA})).
If we
had directly applied the approximation scheme to
the quantum version of $\mathcal{H}_{\bot}$ within $(i)$, the second
term in (\ref{HJ}) would be absent.
One can interpret this term as originally
belonging to the momentum constraints $\mathcal{H}_i$. 

In the bosonic version of the semiclassical approximation
one starts with the constraints as they are derived from the action.
This is why there a term analogously to term (iv) in (\ref{Hbot})
is absent from the very beginning. Therefore, to facilitate the
comparison with the bosonic case and to concentrate on the intrinsic
differences, we shall not take into account term (iv),
\begin{equation}
\label{stoerender term in G^-1}
  {i\hbar n^{AA'}\,\threesD}_i\fal{}{e^{AA'}_i}\,,
\end{equation}
in the following. Nevertheless,
if one wanted, one could carry term (iv) through all the following
expressions; this would, however, not have any consequences for the
main results.  

Since the Hamilton--Jacobi equation of pure general relativity,
Eq. (\ref{bos. HJ eqN.}),
can be recovered in this way, one may try to decompose the full equation
(\ref{susy HJ eq.})
into a part depending only on the tetrad and
a mixed part. We impose in addition the requirement that
we must find the standard classical spacetime background in our
approximation. Therefore we make the ansatz,
\begin{equation}
  S_0[e,\psi]=B_0[e]+F_0[e,\psi]\ ,
  \label{NEW1}
\end{equation}
for the lowest order in the expansion (\ref{NEW21}). On the level of the WKB
wave functional, this corresponds to the
factorization
\begin{displaymath}
  \Psi[e,\psi]=\exp\left(\frac{i}{\hbar}B_0G^{-1}\right)
  \exp\left(\frac{i}{\hbar}(F_0G^{-1}+S_1+...)\right).
\end{displaymath}
The pure bosonic part $B_0$ can be {\em chosen} such that
\begin{equation}
\label{decomp. HJ eq., first part}
  4\pi i n^{AA'}{D^{{BB'}}_{ij}}\fal{B_{0}}{e^{AB'}_j}\fal{B_{0}}{e^{BA'}_i}+ V=0\ ,
\end{equation}
corresponding to the Hamilton--Jacobi equation (\ref{bos. HJ eqN.}).
A solution $B_0$ then determines the condition for the part $F_0$,
\begin{eqnarray}
  \label{decomp. HJ eq., second part}
    0 &=&  4\pi i\Bigg(\psi^B_i\fal{F_0}{e^{AB'}_j}{E^{CAB'i}_{Bjk}}\fal{F_0}{\psi^C_k}
    +\psi^B_i\fal{B_0}{e^{AB'}_j}{E^{CAB'i}_{Bjk}}\fal{F_0}{\psi^C_k}
  \nonumber\\&&
    -n^{AA'}{D^{{BB'}}_{ij}}\fal{F_0}{e^{AB'}_j}\fal{F_0}{e^{BA'}_i}
    -n^{AA'}{D^{{BB'}}_{ij}}\fal{F_0}{e^{AB'}_j}\fal{B_0}{e^{BA'}_i}
    -n^{AA'}{D^{{BB'}}_{ij}}\fal{B_0}{e^{AB'}_j}\fal{F_0}{e^{BA'}_i}\Bigg)
  \nonumber\\
  &&+\frac{i}{\sqrt{h}}\eps^{ijk}e^{BC'}_ie^{A}_{\hspace{1.6mm} C'l}
    \left({\threesD}_j\psi_{Ak}\right)\fal{F_0}{\psi^B_l}\ ,
\end{eqnarray}
which is automatically fulfilled if $S_0$ is a solution of
 (\ref{susy HJ eq.}) -- without the omitted term
(\ref{stoerender term in G^-1}) --, and $F_0=S_0-B_0$.

It is now appropriate to
interpret the solutions of the
Hamilton--Jacobi equation (\ref{susy HJ eq.}). A
particular aspect, distinguishing
this equation from its bosonic analogue (see Appendix B)
is the presence of the gravitino in the
first and third terms. This means that it will generically be present in $S_0$
(or in $F_0$) -- see Eq. (\ref{NEW1}) above.
Moreover, one can indirectly prove that $S_0$ {\em must}
depend on the gravitino. The argument goes as follows.

It is known from the full theory that a pure bosonic
solution, $\Psi[e]$, to the full set of constraints
cannot exist \cite{Carroll:Friedmann:94}.
In fact, this argument can be extended in a
straightforward way to each term in the semiclassical expansion,
as we shall show now.
In analogy to the full theory we act with
the Hermitian conjugated SUSY constraint on $\Psi$
(see Appendix A and in particular Eq. (\ref{constraint, quantized susy}))
and multiply it
with $[\Psi]^{-1}$. With the ansatz
$\Psi=\exp(i[S_0G^{-1}+S_1+S_2G+...]/\hbar)$ we obtain in the lowest order $G^0$:
\begin{eqnarray}
\label{nonbos2}
  [\Psi]^{-1}\bar{S}_{A'}\Psi\stackrel{\mathcal{O}(G^{0})}{=}
  \eps^{ijk}e_{AA'i}\threesD_j\psi^{A}_k
+4\pi i\psi^{A}_i\fal{S_0}{e^{AA'}}_i=0\ .
\end{eqnarray}
Similar to the full theory this must hold for arbitrary fields $\psi^{A}_i$ and $e^{AA'}_i$. First we find
that (\ref{nonbos2}) does not allow trivial solutions, that is, $S_0$ must at least depend on $e^{AA'}_i$.
Otherwise we would get the condition
\begin{displaymath}
  \eps^{ijk}e_{AA'i}\threesD_j\psi^{A}_k=0\ ,
\end{displaymath}
which cannot hold for all fields.
Let us now assume that $S_0$ does not depend on the gravitino field $\psi^{A}_i$. Integrating (\ref{nonbos2}) with
an arbitrary continuous spinorial test function $\bar{\eps}^{A'}(x)$
over space leads to
\begin{displaymath}
  I_0\equiv \int d^3x\bar{\eps}^{A'}\left(\eps^{ijk}e_{AA'i}\threesD_j\psi^{A}_k
  +4\pi i\psi^{A}_i\fal{S_0}{e^{AA'}_i}\right)
  =0\ .
\end{displaymath}
Using as in \cite{Carroll:Friedmann:94}
the replacement ${\psi^{A}_i} \mapsto\psi^{A}_i\exp(\phi(x))$ and $\bar{\eps}^{A'}(x) \mapsto
\bar{\eps}^{A'}(x)\exp(-\phi(x))$, respectively, this yields a new integral,
\begin{eqnarray}
  I_0'&\equiv&\int d^3x\bar{\eps}^{A'}\exp(-\phi)
  \nonumber\\ &&
  \times\left(\eps^{ijk}e_{AA'i}\threesD_j(\exp(\phi)\psi^{A}_k)
  +4\pi i\exp(\phi)\psi^{A}_i\fal{S_0}{e^{AA'}_i}\right)
  =0\ .\nonumber
\end{eqnarray}
 From $\Delta I_0= I_0-I_0'=0$ one gets the same contradiction as in the
the full theory, since
\begin{displaymath}
  \Delta I_0=\int d^3x\,\eps^{ijk}e^{AA'}_i(x)\bar{\eps}^{A'}(x)\psi_{Ak}(x)\partial_j\phi(x)=0
\end{displaymath}
cannot hold for all fields.
In higher orders, the calculation turns out to be simpler than in the lowest order. For $n\geq 1$ we obtain
\begin{equation}
\label{nonbos3}
  [\Psi]^{-1}\bar{S}_{A'}\Psi\stackrel{\mathcal{O}(G^{n})}{=}-4\pi i G^n \psi^{A}_i\fal{S_n}{e^{AA'}}_i=0\ .
\end{equation}
There are two possible conclusions. The first possibility is to assume that $S_n$ does not depend on the
bosonic field $e^{AA'}_i$. This would be very restrictive and, moreover, no proper bosonic limit would
exist. We therefore dismiss this option as irrelevant for the
semiclassical approximation.
The second possibility to satisfy (\ref{nonbos3}) is
to introduce a dependence on the gravitino field at each order.
Hence we must have $S^n\equiv S^n[e,\psi]$ for all $n$.
The consequence
is that the Hamilton--Jacobi equation -- and
therefore the now 
retrieved `background spacetime' -- must necessarily involve the gravitino,
a conclusion identical to what followed
from the SUSY Hamilton--Jacobi equation (\ref{susy HJ eq.}),
or (\ref{decomp. HJ eq., first part}) and (\ref{decomp. HJ eq., second part}).

Let us now elaborate more on this important
feature. As shown in \cite{DeWitt:67} and \cite{Gerlach:69}
(cf. also \cite{OUP} and Appendix B)
for the pure bosonic case
(that is, for pure general relativity),
a solution of the Hamilton--Jacobi equation conveys a
classical spacetime which can serve as the appropriate background for
the higher orders. This is due to the
fact that such a solution is equivalent
to the field equations originating from the
Einstein--Hilbert action. This constitutes
DeWitt's interpretation \cite{DeWitt:67}:
Every solution $S_0$ describes a family of solutions
to the classical field equations. For
every three-geometry there is one member of this family
with a spacelike hypersurface being equal to this three-geometry.
But, as mentioned, the situation
in the semiclassical approximation of
supersymmetric quantum gravity has a particular
difference: {\em the presence of the gravitino}.

In order to address this feature it may
prove relevant to mention the following.
The use of strictly bosonic backgrounds constitutes
the sole procedure in general relativity
and has also been the norm  when dealing  with
classical black hole solutions in
SUGRA and superstring theories \cite{SSt,SUSY}. Being more specific,
it is required that those backgrounds, while
satisfying the equations of motion, be invariant
under SUSY transformations. This leads to conditions,
namely that the
parameters of the SUSY transformation must satisfy a Killing spinor
equation. Nevertheless, there have been some notable
exceptions, see, for example, \cite{AG}--\cite{AG3} and in particular
\cite{AG4}. In \cite{AG} an exact,
asymptotically flat, stationary solution of the
field equations of a SUGRA theory was found,
constituting a supersymmetric generalization of a
black-hole geometry. Subsequently, another type of
solutions describing superpartners to the
bosonic configurations was presented in
\cite{AG1}. In these solutions, the role of the
classical configuration (e.g., the black hole)
is played by a solution with certain fermionic (i.e., gravitino)
field excitations. The full metric solution consists
of a supermultiplet, formed by supertranslated partners to
the purely bosonic configuration (see \cite{AG1} for more
details).

It is in this context that we can interpretate our results
for the SUSY Hamilton--Jacobi equation (\ref{susy HJ eq.}), inducing
a spacetime background with both tetrad (graviton) and fermionic
(gravitino) terms. Being more concrete, such a supersymmetric configuration
will be a solution of the equations of motion
of the theory, with a metric being written as
\begin{equation}
g = g_{\rm B} + g_{\rm S}\ ,
\label{NEW2}
\end{equation}
where the term $g_{\rm B}$ denotes the `body' and
$g_{\rm S}$ the `soul',
adopting the definitions and nomenclature
introduced by DeWitt in \cite{AG4}.
It should be noticed that $g_{\rm B}$ and $g_{\rm S}$
 correspond, respectively, to the purely bosonic and fermionic
sectors.
In other words, a spacetime configuration induced
from a solution of  (\ref{susy HJ eq.}) will
constitute a Grassmann-algebra-valued
field that can be decomposed into the
`body' which takes values in the domain of real or
complex numbers and a `soul' which is
nilpotent \cite{AG4}. Moreover, we take the point of view
that the `body' of the Grassmann-valued field
must be given an operational interpretation and identified
with a standard classical bosonic configuration.
This overall description
corresponds to the scenario of
supermanifolds (and therefore superRiemannian geometries)
thoroughly described in \cite{AG4}.\footnote{A related
 discussion is made in
\cite{AG3}, with the introduction of a line element
 $ds^2 = g_{AB}dz^A dz^B$, where
$z^A=(x^\mu, \theta^\alpha)$, and $\theta^\alpha$ being
Grassmannian coordinates. The metric can then be divided in sectors
such as, e.g, Bose--Bose $g_{\mu\nu}$, and Fermi--Fermi
 $g_{\alpha\beta}$.}

A solution of
 the SUSY Hamilton--Jacobi equation (\ref{susy HJ eq.})
 will thus correspond to a 
 spacetime (yielding an
appropriate background for the higher orders), 
whose metric includes the standard classical 
bosonic sector plus corrections in the form of 
gravitino terms.
Therefore, DeWitt's interpretation could again
be employed: Every such 
solution $S_0$ describes a family of solutions
to the classical field equations. For
every three-geometry there is one member of this
family with an appropriate spacelike hypersurface.
The important additional feature is that
we will then be dealing with a
configuration defined on the space of all
possible spatial tetrads and gravitino fields,
$SuperRiem~\Sigma$.
The standard classical background for Eq.
 (\ref{susy HJ eq.}) and the expansion
 (\ref{NEW1}) can be interpreted as follows:
  $B_0$ together with (\ref{decomp. HJ eq., first part})
and a condition obtained from the expansion of
 the other constraints yields a 
 standard
 classical spacetime without gravitino.
The part $F_0$ and (\ref{decomp. HJ eq., second part}) would then provide
corrections to this. 
Such an interpretation, however, does not close the 
discussion on the issue and further
analysis is certainly required. Finally, and although
perhaps surprising at first glance, the presence of the gravitino
(even at 
higher orders of approximation) is not necessarily in
conflict with observation. Long ago, Pauli has performed a WKB
approximation for a Dirac electron, which has some similarities to the
present scheme \cite{Pauli}: 
The semiclassical approximation of the Dirac equation leads, in the
leading order, to a Hamilton--Jacobi equation for a spinless classical
relativistic particle (where the mass is given by the electron mass).
Only the next order (order $\hbar$)
contains information about the electron spin. 
 In the same way one
might expect that the spin--3/2 nature of the gravitino does not play a
role at the leading order of our semiclassical expansion scheme, that is, 
at the order of the Hamilton--Jacobi equation, 
but that it comes into play only at the following orders.

%%%%%%%%%%%%%%%%%%%%%%%%%%%%%%%%%%%%%%%%%%%%%%%%%%%%%%%%%%%%%%%%%%%%%%%%%%%%%

\section{Recovery of the functional Schr\"odinger equation\label{Schroedinger}}

We shall now proceed with the semiclassical expansion of
(\ref{Hbot}), (\ref{Hamiltonian constraint, formal}).
At order $G^{0}$ we expect to recover
the functional Schr\"odinger equation.
Neglecting the contribution of term (iv) in (\ref{Hbot}),
we find
 \begin{eqnarray}
  \label{contributions in G^0 complete}
    0 & =&\frac{1}{2}\left(\frac{1}{\sqrt{h}}\fal{^2S_1}{\Phi^2}
  -\frac{i\hbar}{\sqrt{h}}\fal{S_1}{\Phi}\fal{S_1}{\Phi}
  +\sqrt{h}h^{ij}\Phi_{,i}\Phi_{,j}+\sqrt{h}(m^2\Phi^2+U(\Phi))\right)
  \nonumber\\&&
    +4\pi\Bigg[2i \psi^B_i\fal{S_{(0}}{e^{AB'}_j}{E^{CAB'i}_{Bjk}}\fal{S_{1)}}{\psi^C_k}
    -\hbar \psi^B_i{E^{CAB'i}_{Bjk}}\zfal{S_{0}}{e^{AB'}_j}{\psi^C_k}
  \nonumber\\&&
    -\hbar\left(\frac{3i}{\sqrt{h}}\psi^C_k
    +\psi^{Bj}\eps_{jkl}n^{CB'}e^l_{BB'}\right)\fal{S_1}{\psi^C_k}
    -2 i n^{AA'}\fal{S_{(0}}{e^{AB'}_j} {D^{{BB'}}_{ij}}\fal{S_{1)}}{e^{BA'}_i}
  \nonumber\\&&
    -\frac{i\hbar}{\sqrt{h}}e^{BA'}_i\fal{S_0}{e^{BA'}_i}
    -\hbar n^{AA'}{D^{{BB'}}_{ij}}\zfal{S_{0}}{e^{AB'}_j}{e^{BA'}_i}\Bigg]
  \nonumber\\&&
    +\frac{i}{\sqrt{h}}\eps^{ijk}e^{BC'}_ie^{A}_{\hspace{1.6mm} C'l}
    \left({\threesD}_j\psi_{Ak}\right)\fal{S_1}{\psi^B_l}\ .
\end{eqnarray}
In analogy to Appendix \ref{Scheme} we simplify this equation
by introducing the wave functional
\begin{equation}
\label{susy wave functional chi}
  \chi=W[e,\psi]\exp\left(\frac{i}{\hbar}S_1[e,\psi,\Phi]\right)\ ,
\end{equation}
cf. Eq. (\ref{wave functional chi}),
by demanding the following condition for the WKB prefactor $W$,
\begin{eqnarray}
\label{condition for W}
    0 & = & \psi^B_i\fal{S_0}{e^{AB'}_j}{E^{CAB'i}_{Bjk}}\fal{W}{\psi^C_k}
    +\psi^B_i\fal{W}{e^{AB'}_j}{E^{CAB'i}_{Bjk}}\fal{S_0}{\psi^C_k}
  \nonumber\\&&
    -n^{AA'}\fal{S_0}{e^{AB'}_j} {D^{{BB'}}_{ij}}\fal{W}{e^{BA'}_i}
    -n^{AA'}\fal{W}{e^{AB'}_j} {D^{{BB'}}_{ij}}\fal{S_0}{e^{BA'}_i}
  \nonumber\\&&
    +\frac{1}{4\pi\sqrt{h}}\eps^{ijk}e^{BC'}_ie^{A}_{\hspace{1.6mm} C'l}
    \left({\threesD}_j\psi_{Ak}\right)\fal{W}{\psi^B_l}
  \nonumber\\&&
    -\Bigg[\psi^B_i{E^{CAB'i}_{Bjk}}\zfal{S_{0}}{e^{AB'}_j}{\psi^C_k}
    +\left(\frac{3i}{\sqrt{h}}\psi^C_k+\psi^{Bj}\eps_{jkl}n^{CB'}e^l_{BB'}\right)\fal{S_0}{\psi^C_k}
  \nonumber\\&&
    +n^{AA'}{D^{{BB'}}_{ij}}\zfal{S_{0}}{e^{AB'}_j}{e^{BA'}_i}
    +\frac{i}{\sqrt{h}}e^{BA'}_i\fal{S_{0}}{e^{BA'}_i}\Bigg]W\ .
\end{eqnarray}
We can then rewrite this condition in the form
\begin{eqnarray}
\label{con.-law for W}
    & & n^{AA'}\fal{}{e^{AB'}_j}\left({D^{{BB'}}_{ij}}\fal{S_0}{e^{BA'}_i}W\
    -\psi^B_i\eps^{ilm}\DD\fal{S_0}{\psi^C_k}W\right)
  \nonumber\\
   \;  =& & n^{AA'}\fal{S_0}{e^{AB'}_j} {D^{{BB'}}_{ij}}\fal{W}{e^{BA'}_i}
    -\psi^B_i\fal{S_0}{e^{AB'}_j}{E^{CAB'i}_{Bjk}}\fal{W}{\psi^C_k}
  \nonumber\\
    &&-\left(\frac{3i}{\sqrt{h}}\psi^C_k+\psi^{Bj}\eps_{jkl}n^{CB'}e^l_{BB'}\right)\fal{S_0}{\psi^C_k}\ .
\end{eqnarray}
We recognize that the
right-hand side of Eq. (\ref{con.-law for W}) prevents it 
to be interpreted  as a conservation law
(see Eq. (\ref{conservation law})) in the 
context of $SuperRiem~\Sigma$. 
As expected, only in the very special case of a vanishing dependence 
of $S_0$ and $W$ on the gravitino can a conservation law be formulated. 
By assuming 
that $S_0[e]$ and $W[e]$
can be rewritten as $S_0[h_{ij}]$ and $W[h_{ij}]$, 
we then obtain a simpler expression from 
(\ref{con.-law for W}), in the form of a 
conservation equation, 
\begin{equation}
\label{simpler form}
  n^{AA'}\fal{}{e^{AB'}_j}\left(\DBij\fal{S_0}{e^{BA'}_i}W^{-2}\right)=0\,.
\end{equation}
However, as we have seen above, $S_0$ \emph{must} depend on the gravitino. 

 Inserting (\ref{susy wave functional chi}) and (\ref{condition for W}) into
(\ref{contributions in G^0 complete}), we find the Tomonaga--Schwinger
equation,
\begin{eqnarray}
  \label{susy Schroedinger eq.}
    &&\hspace{-10mm} 4\pi i(i\hbar) \Bigg[\psi^B_i\fal{S_{0}}{e^{AB'}_j}{E^{CAB'i}_{Bjk}}\fal{}{\psi^C_k}
    +\psi^B_i{E^{CAB'i}_{Bjk}}\fal{S_{0}}{\psi^C_k}\fal{}{e^{AB'}_j}
  \nonumber\\&&
    -n^{AA'}\fal{S_0}{e^{AB'}_j} {D^{{BB'}}_{ij}}\fal{}{e^{BA'}_i}
    -n^{AA'}\fal{S_0}{e^{BA'}_i}{D^{{BB'}}_{ij}}\fal{}{e^{AB'}_j}
  \nonumber\\&&
    +\frac{1}{4\pi\sqrt{h}}\eps^{ijk}e^{BC'}_ie^{A}_{\hspace{1.6mm} C'l}
    \left({\threesD}_j\psi_{Ak}\right)\fal{}{\psi^B_l}\Bigg]\chi
    \equiv i\hbar\fal{\chi}{\tau}=\Hmbot\chi\ .
\end{eqnarray}
The time functional $\tau(x;e,\psi]$ is defined by
\begin{eqnarray}
  \label{susy time functional}
    &&\hspace{-10mm} 4\pi\Bigg[\psi^B_i\fal{S_{0}}{e^{AB'}_j(y)}{E^{CAB'i}_{Bjk}}\fal{}{\psi^C_k(y)}
    +\psi^B_i{E^{CAB'i}_{Bjk}}\fal{S_{0}}{\psi^C_k(y)}\fal{}{e^{AB'}_j(y)}
  \nonumber\\&&
    -n^{AA'}\fal{S_0}{e^{AB'}_j(y)} {D^{{BB'}}_{ij}}\fal{}{e^{BA'}_i(y)}
    -n^{AA'}\fal{S_0}{e^{BA'}_i(y)}{D^{{BB'}}_{ij}}\fal{}{e^{AB'}_j(y)}
  \nonumber\\&&
    +\frac{1}{4\pi\sqrt{h}}\eps^{ijk}e^{BC'}_ie^{A}_{\hspace{1.6mm} C'l}
    \left({\threesD}_j\psi_{Ak}\right)\fal{}{\psi^B_l(y)}\Bigg]\tau(x;e,\psi]
  =\delta(x-y)\ .
\end{eqnarray}
For clarity we show the arguments $(y)$ of the functional derivatives on the left-hand
side. Note that, of course, all quantities involving the tetrad or the gravitino on this side depend on $y$. The functional Schr\"odinger equation
is found from (\ref{susy Schroedinger eq.}) after integration over space.

One may wish to separate (\ref{susy Schroedinger eq.})
into a bosonic and a fermionic part,
in analogy to the treatment of (\ref{susy HJ eq.}).
In addition to the already decomposed $S_0$ we try the ansatz,
\begin{equation}
\label{decomp. of S1}
  S_1=B_1[e,\Phi]+F_1[e,\psi,\Phi]
\end{equation}
and assume a product ansatz for the WKB prefactor,
\begin{displaymath}
  W[e,\psi]=W^b[e]W^f[\psi]\ .
\end{displaymath}
The wave functional $\chi$ can be factorized as
\begin{equation}
  \chi=\tilde{\chi}\xi\ ,
\end{equation}
where
\begin{equation}
  \tilde{\chi}=W^b\exp\left(\frac{i}{\hbar}B_1\right), \quad
  \xi=W^f\left(\frac{i}{\hbar}F_1\right).
\end{equation}
Now we see that an expansion of (\ref{susy Schroedinger eq.})
with decomposed $S_0$ and $S_1$ contains a part of the form
\begin{eqnarray}
\label{mixed Schroedinger terms}
 4\pi i(i\hbar) \Bigg[\psi^B_i{E^{CAB'i}_{Bjk}}\fal{F_{0}}{\psi^C_k}\fal{}{e^{AB'}_j}
    -n^{AA'}\fal{F_0}{e^{AB'}_j} {D^{{BB'}}_{ij}}\fal{}{e^{BA'}_i}
    -n^{AA'}\fal{F_0}{e^{BA'}_i}{D^{{BB'}}_{ij}}\fal{}{e^{AB'}_j}\Bigg]\tilde{\chi}\ .\hspace{4mm}
\end{eqnarray}
To demand that these terms vanish is {\em not} possible, since $F_0$ is already determined by the Hamilton--Jacobi
equation (\ref{susy HJ eq.}) and $\tilde{\chi}$ should be a solution of the reduced local Schr\"odinger
equation (\ref{Schroedinger eq., second version}), see below. This
suggests that the requirement for a local Schr\"odinger
equation that does not depend on the gravitino is impossible to achieve.
We could of course simply demand that
\begin{eqnarray}
\label{Schroedinger eq., second version}
   & &\hspace{-1cm}-4\pi (i\hbar)i\Bigg[
    \fal{S_{0}}{e^{AB'}_j} n^{AA'}{D^{{BB'}}_{ij}}\fal{}{e^{BA'}_i}
    +n^{AA'}{D^{{BB'}}_{ij}}\fal{S_{0}}{e^{BA'}_i}\fal{}{e^{AB'}_j}
\Bigg]\tilde{\chi}\equiv i\hbar\fal{\tilde{\chi}}{\tilde{\tau}}=\Hmbot\tilde{\chi}\hspace{2mm}
\end{eqnarray}
holds with a redefined time functional $\tilde{\tau}$.
But in order to obtain this, we have to impose various
additional conditions, namely for the factors $W^b$ and
$W^f$, as well as for the part $F_1$ that depends on
the solution $\tilde{\chi}$ of (\ref{Schroedinger eq.,
second version}). In particular, the part $F_1$ would
be determined by the matter field, since (\ref{mixed
Schroedinger terms}) does not vanish in general. This
should be carefully analysed in view of the
 ansatz (\ref{decomp. of S1}).
Furthermore, additional conditions
are hard to justify and may be without any physical meaning. The lesson learnt
from this is that the presence of the gravitino is mandatory for the definition
of the time functional as well as for the Schr\"odinger equation.

Nevertheless, the interpretation of the time functional should
be similar to the one given in Appendix
\ref{Scheme}: It defines a local (`many-fingered') time parameter.
However, this should now be
on the space of all possible spatial tetrad
and gravitino fields, $SuperRiem~\Sigma$. The question of how to
interpret a classical background containing the gravitino,
inducing this type of time functional, would require the
discussion and proposed interpretation presented in the last section.

Finally, let us indicate that the
functional Schr\"odinger equation can only be recovered in this way
if a real solution $S_0$ to the Hamilton--Jacobi equation is chosen.
 One would not have been able to derive it from, for example,
a superposition $\propto (\exp(iS_0)+\exp(-iS_0))$. This problem arises,
of course, already in the non-supersymmetric case where it was shown that
the components in such a superposition become effectively independent
due to decoherence by additional degrees of freedom \cite{deco}.
The same is expected to hold here. Decoherence should be efficient
during the greatest part of the evolution of the universe. In some
regions (such as the Planck regime or the region corresponding to
a classical turning point) the various semiclassical components may
interfere with each other and thereby spoil the validity of the
approximation scheme presented here \cite{CK88}.

%%%%%%%%%%%%%%%%%%%%%%%%%%%%%%%%%%%%%%%%%%%%%%%%%%%%%%%%%%%%%%%%%%%%%%%%%%%%

\section{Corrections to the Schr\"odinger equation\label{Corrections}}

We shall now continue with the semiclassical expansion scheme.
At the order $G^1$ we find the following equation:
\begin{eqnarray}
\label{contributions in G^1}
    \Psi^{-1}(H_{\bot}+\Hmbot) \Psi &\stackrel{\mathcal{O}(G^{1})}{=}&
    4\pi{G}\Bigg[\underbrace{i\psi^B_i\fal{S_{1}}{e^{AB'}_j}\E\fal{S_{1}}{\psi^C_k}}_{\rm{(i)}}
    +\underbrace{\hbar \psi^B_i\E\zfal{S_{1}}{e^{AB'}_j}{\psi^C_k}}_{\rm{(ii)}}
  \nonumber\\ &&
    -\underbrace{\hbar\left(\frac{3i}{\sqrt{h}}\psi^C_k
    +\psi^{Bj}\eps_{jkl}n^{CB'}e^l_{BB'}\right)\fal{S_1}{\psi^C_k}}_{\rm{(iii)}}
  \nonumber\\ &&
    +\underbrace{2i\psi^B_i\fal{S_{(0}}{e^{AB'}_j}\E\fal{S_{2)}}{\psi^C_k}}_{\rm{(iv)}}
    -\underbrace{2i n^{AA'}\fal{S_{(0}}{e^{AB'}_j} \DBij\fal{S_{2)}}{e^{BA'}_i}}_{\rm{(v)}}
  \nonumber\\ &&
    -\underbrace{i n^{AA'}\fal{S_{1}}{e^{AB'}_j} \DBij\fal{S_{1}}{e^{BA'}_i}}_{\rm{(vi)}}
    -\underbrace{\hbar n^{AA'}\DBij\zfal{S_{1}}{e^{AB'}_j}{e^{BA'}_i}}_{\rm{(vii)}}
  \nonumber\\ &&
    -\underbrace{\frac{i\hbar}{\sqrt{h}}e^{BA'}_i\fal{S_1}{e^{BA'}_i}}_{\rm{(viii)}}\Bigg]
    +\underbrace{\frac{iG}{\sqrt{h}}\eps^{ijk}e^{BC'}_ie^{A}_{\hspace{1.6mm} C'l}
    \left({\threesD}_j\psi_{Ak}\right)\fal{S_2}{\psi^B_l}}_{\rm{(ix)}}
  \nonumber\\ &&
    +\underbrace{\frac{G}{2\sqrt{h}}
    \Bigg[2\fal{S_1}{\Phi}\fal{S_2}{\Phi}-i\hbar\fal{^2S_2}{\Phi^2}}_{\rm{(x)}}\Bigg]=0\ .
 \end{eqnarray}
In order to obtain the Schr\"odinger equation with corrections of order $G$, we must perform various steps on a
formal level, which are very similar to those applied in \cite{KS91}. The derivation is straightforward but
lengthy. We therefore relegate some of the calculations to
 Appendix D and present here only the results and
their physical discussion.

First we use equations (\ref{A1})--(\ref{A8}) to rewrite the expressions containing $S_1$ in
(\ref{contributions in G^1}) in terms of $\chi$ and $W$. We also use the definition (\ref{susy time
functional}) of the time functional $\tau(x;e,\psi]$ and make the decomposition
\begin{equation}
  S_2[e,\psi,\Phi]=\sigma_2[e,\psi]+\eta[e,\psi,\Phi]
\end{equation}
in order to separate the pure gravitational parts of (\ref{contributions in G^1}) from those containing the
matter field \cite{KS91}.
By demanding the following condition for $\sigma_2[e,\psi]$,
\begin{eqnarray}
\label{cond. for sigma2}
    \fal{\sigma_2}{\tau}&=&
    -\frac{4\pi\hbar^2}{iW}\Bigg[\psi^B_i\E\left(\frac{2}{W}\fal{W}{{e^{AB'}_j}}\fal{W}{\psi^C_k}
    -\zfal{W}{{e^{AB'}_j}}{\psi^C_k}\right)
  \nonumber\\ &&
    +\left(\frac{3i}{\sqrt{h}}\psi^C_k
    +\psi^{Bj}\eps_{jkl}n^{CB'}e^l_{BB'}\right)\fal{W}{\psi^C_k}
  \nonumber\\ &&
    -\frac{2}{W}n^{AA'}\DBij\fal{W}{{e^{AB'}_j}}\fal{W}{{e^{BA'}_i}}+n^{AA'}\DBij\zfal{W}{{e^{AB'}_j}}{{e^{BA'}_i}}
    -\frac{i}{\sqrt{h}}e^{BA'}_i\fal{W}{{e^{BA'}_i}}\Bigg],
\end{eqnarray}
which is motivated by the analogous step in the
standard quantum mechanical WKB expansion
 \cite{LL}, we can rewrite
(\ref{contributions in G^1}) as
\begin{eqnarray}
  \label{contribution in G^1, second version}
    \fal{\eta}{\tau} &=&
    -\frac{4\pi\hbar^2}{i\chi}\Bigg[\psi^B_i
    \E\Bigg(\zfal{\chi}{{e^{AB'}_j}}{\psi^C_k}
    -\frac{1}{W}\left\{\fal{W}{{e^{AB'}_j}}\fal{\chi}{\psi^C_k}+\fal{\chi}{{e^{AB'}_j}}\fal{W}{\psi^C_k}\right\}\Bigg)
  \nonumber\\ &&
    -\left(\frac{3i}{\sqrt{h}}\psi^C_k
    +\psi^{Bj}\eps_{jkl}n^{CB'}e^l_{BB'}\right)\fal{\chi}{\psi^C_k}\Bigg]
  \nonumber\\ &&
    +\frac{4\pi\hbar^2}{i\chi}\Bigg[
    n^{AA'}\DBij\zfal{\chi}{{e^{AB'}_j}}{{e^{BA'}_i}}
    -\frac{1}{W}n^{AA'}\DBij\left\{\fal{W}{{e^{AB'}_j}}\fal{\chi}{{e^{BA'}_i}}+\fal{\chi}{{e^{AB'}_j}}\fal{W}{{e^{BA'}_i}}\right\}
  \nonumber\\ &&
    +\frac{i}{\sqrt{h}}e^{BA'}_i\fal{\chi}{{e^{BA'}_i}}\Bigg]
    +\frac{iG\hbar }{2\sqrt{h}}\Bigg[\frac{2}{\chi}\fal{\chi}{\Phi}\fal{\eta}{\Phi}
    +\fal{^2\eta}{\Phi^2}\Bigg].
\end{eqnarray}
Up to the current order the wave functional has assumed the form
\begin{eqnarray}
    \Psi & = & \exp\left(\frac{i}{\hbar}\left[S_0 G^{-1}+S_1+S_2G\right]\right)
  \nonumber\\
     &=&  \frac{1}{W}\exp \left(\frac{i}{\hbar}\left[S_0 G^{-1}+\sigma_2 G\right]\right)
        \chi\exp\left(\frac{i}{\hbar}\eta G\right).
\end{eqnarray}
Since we have already fixed the pure gravitational phase $\sigma_2$ in (\ref{cond. for sigma2}), and we are mainly
interested in the matter part, we can restrict our attention to
\begin{equation}
\label{wave functional theta}
  \Theta\equiv\chi\exp\left(\frac{i}{\hbar}\eta G\right),
\end{equation}
which contains the functional $\chi$ and the not yet determined part $\eta$ of $S_2$.  Now we multiply the uncorrected
equation (\ref{susy Schroedinger eq.}) with
$\exp\left(i\eta G/\hbar\right)$ and add it to equation (\ref{contribution in G^1, second version})
multiplied by $-G\chi\exp\left(i\eta G/\hbar\right)$. Using (\ref{A9}) and (\ref{A10}),
we can perform the next steps and obtain the local Schr\"odinger equation with corrections up to the order
$G^1$:
\begin{eqnarray}
\label{corrected Schroedinger eq., 1. version}
    i\hbar\fal{\Theta}{\tau}&=&\Hmbot\Theta
    +\frac{4\pi G\hbar^2}{i\chi}\Bigg[-\frac{1}{W}\psi^B_i\E\
    \left(\fal{W}{{e^{AB'}_j}}\fal{\chi}{\psi^C_k}+\fal{\chi}{{e^{AB'}_j}}\fal{W}{\psi^C_k}\right)
   \nonumber\\ &&
    +\psi^B_i\E\zfal{\chi}{{e^{AB'}_j}}{\psi^C_k}
    -\left(\frac{3i}{\sqrt{h}}\psi^C_k
    +\psi^{Bj}\eps_{jkl}n^{CB'}e^l_{BB'}\right)\fal{\chi}{\psi^C_k}\Bigg]\Theta
  \nonumber\\ &&
    -\frac{4\pi G\hbar^2}{\chi}\Bigg[n^{AA'}\DBij\zfal{\chi}{{e^{AB'}_j}}{{e^{BA'}_i}}-\frac{1}{W}
    n^{AA'}\DBij\left(\fal{W}{{e^{AB'}_j}}\fal{\chi}{{e^{BA'}_i}}+\fal{\chi}{{e^{AB'}_j}}\fal{W}{{e^{BA'}_i}}\right)
  \nonumber\\ &&
    +\frac{i}{\sqrt{h}}e^{BA'}_i\fal{\chi}{{e^{BA'}_i}}\Bigg]\Theta\ .
\end{eqnarray}
On a formal level, the terms in the third line have the same structure as those which already appeared in the
expansion of the Wheeler-DeWitt equation \cite{KS91}.

For further treating (\ref{corrected Schroedinger eq., 1. version}), we apply the idea of the
decomposition into a `normal' and a `tangential' part \cite{KS91},
where `normal' means normal to hypersurfaces $S_0=$ constant (thus being
directed along the classical spacetimes defined by $S_0$), and `tangential'
means tangential to $S_0=$ constant, see below.
 For this purpose it
is appropriate to introduce a metric $\mathcal{G}$ on the
 space $SuperRiem~\Sigma$, which we shall call
`Super-DeWitt metric' and whose properties still have to be
investigated.
 This metric should
be the supersymmetric analogue of the DeWitt metric $G_{ijkl}$. The
main difference to Appendix  \ref{Scheme} is that
we must consider now $SuperRiem~\Sigma$,
 the direct sum of the tetrad space and
the gravitino space. It contains vectors of the form
$(e^{{AA'} i},\psi^B_j)\equiv q_{a} $. Henceforth, the
latin indices starting with $a$ are `condensed'
superindices which run through all bosonic and fermionic
degrees of freedom.

The metric $\mathcal{G}$ reads in block form
\begin{equation}
\label{blockform}
  \mathcal{G}_{ab}=\left(\begin{array}{cc}
    \mathcal{B} & \mathcal{S}_1 \\
    \mathcal{S}_2 & \mathcal{F} \\
  \end{array}\right).
\end{equation}
The blocks $\mathcal{B}$ and $\mathcal{F}$ denote the pure bosonic and pure fermionic part, respectively;
$\mathcal{S}_1$ and $\mathcal{S}_2$ are the mixed off-diagonal parts. We determine the blocks by the
requirement that the metric applied to the vectors $\delta S_0/\delta q_a \equiv (\delta S_0 / \delta
e^{AA'}_i,\delta S_0/\delta\psi^B_j)$ and $\delta\chi/\delta q_b$ yields all terms containing two derivatives
in the local Schr\"odinger equation (\ref{susy Schroedinger eq.}). The explicit form of the blocks can be read
off immediately. For $\mathcal{B}$ one gets
\begin{equation}
  \mathcal{B}=-4\pi i (n^{AA'}D^{BB'}_{ij}+n^{BB'}D^{AA'}_{ji})\ .
\end{equation}
Since (\ref{susy Schroedinger eq.}) contains no terms with a double derivative with respect to $\psi^A_i$,
the lower diagonal block $\mathcal{F}$ vanishes,
\begin{displaymath}
  \mathcal{F}=0\ .
\end{displaymath}
For $\mathcal{S}_1$ and $\mathcal{S}_2$ we get
\begin{eqnarray}
    \mathcal{S}_1=4\pi i n^{BB'}\psi^{C}_j\eps^{jkm}
    D_{C\hspace{1.5mm}mi}^{\hspace{1.5mm}A'}D^D_{\hspace{1.5mm}B'lk}\ , \nonumber\\
    \mathcal{S}_2=4\pi i n^{AA'}\psi^{B}_i\eps^{ilm}\DD\ .
\end{eqnarray}
With arbitrary vectors
\begin{equation}
  v^{a}=\left(\begin{array}{c}
              B_{AB'}^{j} \\
              F_{D}^{l} \\
      \end{array}\right) \quad \textnormal{and}\quad
  \tilde{v}^b=\left(\begin{array}{c}
              \tilde{B}_{BA'}^{i} \\
              \tilde{F}_{C}^{k} \\
      \end{array}\right),
\end{equation}
we then obtain
\begin{eqnarray}
    \mathcal{G}_{ab}v^{a}\tilde{v}^{b}&=&
    -4\pi i (n^{AA'}D^{BB'}_{ij}+n^{BB'}D^{AA'}_{ji})B_{AB'}^{j}\tilde{B}_{BA'}^{i}
  \nonumber\\ &&
    +4\pi i n^{BB'}\psi^{C}_j\eps^{jkm}D_{C\hspace{1.5mm}mi}^{\hspace{1.5mm}A'}D^D_{\hspace{1.5mm}B'lk}
    F_{D}^{l}\tilde{B}_{BA'}^{i}
  \nonumber\\ &&
    +4\pi i n^{AA'}\psi^{B}_i\eps^{ilm}\DD B_{AB'}^{j}\tilde{F}_{C}^{k}\ .
\end{eqnarray}
For reasons of consistency, the upper diagonal part should contain the DeWitt metric. Of course, it cannot be
exactly the DeWitt metric due to the change of the fundamental bosonic field from $h_{ij}$ to $e^{AA'}_i$. In
some sense it is a tetrad version of it, as we can easily see. We just have to apply the transformation
(\ref{APP: trafo from e to h}). Let $a[e]$ and $b[e]$ be two arbitrary functionals that can also be written
as $a[h_{ij}]$ and $b[h_{ij}]$. We then have
\begin{eqnarray}
\label{recovery of the DeWitt metric}
    &&4\pi i\left(n^{AA'}\DBij\fal{a}{e^{AB'}_j}\fal{b}{e^{BA'}_i}
    +n^{BB'} D^{AA'}_{ji}\fal{a}{e^{AB'}_j}\fal{b}{e^{BA'}_i}\right)
  \nonumber\\
    &&\hspace{3cm}=-32\pi G_{iljk}\fal{a}{h_{jk}}\fal{b}{h_{il}}\ .
\end{eqnarray}
Therefore, we see that for quantities on superspace that can be written in terms of the three-metric
$h_{ij}$ and the gravitino the block $\mathcal{B}$ \emph{is}
the DeWitt metric.\footnote{Note that the factor
$32\pi$ arises due to the choice of our expansion parameter. It can be removed by a simple rescaling as
in Section \ref{HamiltonJacobi}. The minus sign appears due to our convention
for the definition of (\ref{blockform}).}

For further abbreviation and a clearer notation we introduce the operator
\begin{equation}
\label{definition of A}
  A:=\frac{i}{\sqrt{h}}\eps^{ijk}e^{BC'}_ie^{A}_{\hspace{1.6mm} C'l}\left({\threesD}_j\psi_{Ak}\right)\fal{}{\psi^B_l}\ .
\end{equation}
Using these definitions, the Hamilton--Jacobi equation (\ref{susy HJ eq.})
without the omitted term (\ref{stoerender term in G^-1})
assumes the condensed form
\begin{equation}
\label{short H.-J. eq.}
  \frac{1}{2}\mathcal{G}_{ab}\fal{S_0}{q_a}\fal{S_0}{q_b}+A(S_0)-V=0\ .
\end{equation}
We also obtain a short form of the local Schr\"odinger equation (\ref{susy Schroedinger eq.}),
\begin{equation}
\label{short S. eq.}
  i\hbar\mathcal{G}_{ab}\fal{S_0}{q_a}\fal{\chi}{q_b}+i\hbar A{\chi}=i\hbar\fal{\chi}{\tau}=\Hmbot\chi\ .
\end{equation}
The corrected local Schr\"odinger equation (\ref{corrected Schroedinger eq., 1. version}) then reads
\begin{eqnarray}
  \label{corrected Schroedinger eq., 2. version}
    i\hbar\fal{\Theta}{\tau}&=
    &\Hmbot\Theta+\frac{G\hbar^2}{\chi}\Bigg[
    \frac{1}{W}\mathcal{G}_{ab}\fal{\chi}{q_a}\fal{W}{q_b}
    -\frac{1}{2}\mathcal{G}_{ab}\zfal{\chi}{q_a}{q_b}
  \nonumber\\ &&
    +4\pi i\left(\frac{3i}{\sqrt{h}}\psi^C_k +\psi^{Bj}\eps_{jkl}n^{CB'}
    e^l_{BB'}\right)\fal{\chi}{\psi^C_k}
    - \frac{4\pi}{\sqrt{h}}e^{BA'}_i\fal{\chi}{e^{BA'}_i}\Bigg]\Theta\ .
\end{eqnarray}
If the potential term $V$ vanished, this would be the mathematical expression with which we would have to
work. In our case of non-vanishing $V$ it makes sense to decompose the correction terms in a normal and a
tangential part, as mentioned above \cite{KS91}.
 The directions
are defined with respect to hypersurfaces $SuperRiem~\Sigma$ in which $S_0=\rm{constant}$ holds. The
normal part is given by a vector parallel to $\delta S_0/\delta q_a$ and the tangential part by a vector
orthogonal to $\delta S_0/\delta q_a$. In other words,
we consider a trajectory of a classical spacetime\footnote{Let us
remind that we use a  notion of `classical spacetime',
as an element of $SuperRiem~\Sigma$, which for
our configuration involves the gravitino (see section II and
discussion at the end). More precisely, our background spacetime could
be interpreted as a  `classical spacetime with 
correction terms involving gravitinos'.}
in configuration space and split it into a part in the
direction of the evolution and a part
transverse to it (see Appendix B and references therein for
more details).

For the decomposition of the first correction term in (\ref{corrected Schroedinger eq., 2. version}),
$\mathcal{G}_{ab}\delta{\chi}/\delta{q_a}\,\delta{W}/\delta{q_b}$, we make the ansatz
\begin{equation}
\label{decomp. of the first cor. term}
  \mathcal{G}_{ab}\fal{\chi}{q_a}=\gamma\mathcal{G}_{ab}\fal{S_0}{q_a}+T_b\ .
\end{equation}
Therein, $\gamma$ denotes a factor which we determine as follows.
For the tangential part $T_b$,
\begin{equation}
  T_b\fal{S_0}{q_b}=0
\end{equation}
holds.

Multiplication of (\ref{decomp. of the first cor. term}) by $\delta S_0/\delta q_b$ yields
\begin{equation}
  \mathcal{G}_{ab}\fal{\chi}{q_a}\fal{S_0}{q_b}=\gamma\mathcal{G}_{ab}\fal{S_0}{q_a}\fal{S_0}{q_b}\ .
\end{equation}
Making use of (\ref{short H.-J. eq.}) and (\ref{short S. eq.}) we get
\begin{equation}
  \gamma=\frac{(\Hmbot-i\hbar A)\chi}{2i\hbar\tilde{V}}\ ,
\end{equation}
where $\tilde{V}=(V-AS_0)$ denotes a modified potential.

The second-derivative terms in (\ref{corrected Schroedinger eq., 2. version}) can be decomposed by
differentiating (\ref{decomp. of the first cor. term}) with respect to $q_b$,
\begin{eqnarray}
\label{decomp. of the second derivative}
  \mathcal{G}_{ab}\zfal{\chi}{q_a}{q_b}&=&
    -\fal{\mathcal{G}_{ab}}{q_b}\fal{\chi}{q_a}
    +\fal{T_a}{q_a}
    +\fal{\gamma}{q_a}\mathcal{G}_{ab}\fal{S_0}{q_a}
    +\gamma\mathcal{G}_{ab}\zfal{S_0}{q_a}{q_b}
    +\gamma\fal{\mathcal{G}_{ab}}{q_b}\fal{S_0}{q_a}
  \nonumber\\
    &=&
    \fal{\gamma}{q_a}\mathcal{G}_{ab}\fal{S_0}{q_a}
    +\gamma\mathcal{G}_{ab}\zfal{S_0}{q_a}{q_b}
    +\tilde{T},
\end{eqnarray}
where
$\tilde{T}$ denotes the sum of the tangential parts. We now rewrite the condition (\ref{condition for W}) in
terms of the metric $\mathcal{G}_{ab}$ and the operator $A$,
\begin{eqnarray}
\label{short full condition for W}
    0&=&\mathcal{G}_{ab}\fal{S_0}{q_a}\fal{W}{q_b}-A(W)-\frac{W}{2}\mathcal{G}_{ab}\zfal{S_0}{q_a}{q_b}
  \nonumber\\ &&
    -4\pi iW\left(\frac{i}{\sqrt{h}}e^{BA'}_i\fal{S_0}{e^{BA'}_i}
    +\left(\frac{3i}{\sqrt{h}}\psi^C_k
    +\psi^{Bj}\eps_{jkl}n^{CB'}e^l_{BB'}\right)\fal{S_0}{\psi^C_k}\right).
\end{eqnarray}
Using (\ref{decomp. of the first cor. term}), (\ref{decomp. of the second derivative}), and (\ref{short full
condition for W}), we can write the correction terms in the form
\begin{eqnarray}
\label{decomp of the corrections in C_n and C_t}
    &&\hspace{-1cm}\frac{G\hbar^2}{\chi}\Bigg[\frac{1}{W}\mathcal{G}_{ab}\fal{\chi}{q_a}\fal{W}{q_b}
    -\frac{1}{2}\mathcal{G}_{ab}\zfal{\chi}{q_a}{q_b}-\frac{4\pi}{\sqrt{h}}e^{BA'}_i\fal{\chi}{e^{BA'}_i}
  \nonumber\\ &&\hspace{-1cm}
    +4\pi i\left(\frac{3i}{\sqrt{h}}\psi^C_k
    +\psi^{Bj}\eps_{jkl}n^{CB'}e^l_{BB'}\right)\fal{\chi}{\psi^C_k}\Bigg]\Theta=C_n+C_t\ .
\end{eqnarray}
We do not consider the tangential part $C_t$ any further. It was discussed
in the non-supersymmetric case in \cite{BK},
where technical and
physical interpretations can be found. Due to
the complicated formalism of supergravity, we restrict
ourselves to the normal part $C_n$, which in analogy to the bosonic case
is expected anyway to contain the dominating terms.
 To obtain an explicit form of
 it, we need a decomposition of the third and fourth term
on the left-hand side of (\ref{decomp of the corrections in C_n and
 C_t}). It is obtained by defining
\begin{eqnarray}
  w_a:=\left(\frac{i}{\sqrt{h}}e^{BA'}_i,\frac{3i}{\sqrt{h}}\psi^C_k
    +\psi^{Bj}\eps_{jkl}n^{CB'}e^l_{BB'}\right)
\end{eqnarray}
and writing
\begin{equation}
  w_a\fal{\chi}{q_a}=\gamma w_a\fal{S_0}{q_a}+w_a\tilde{T}^a\ ,
\end{equation}
where $\tilde{T}^a$ is the tangential part. Making use of all preparations, we find that in the normal part
many terms cancel out, and we obtain the form
\begin{eqnarray}
\label{normal part with chi and theta}
      C_n&=&\frac{G\hbar^2}{\chi}
      \Bigg[-\gamma\frac{(AW)}{W}-\frac{1}{2}\mathcal{G}_{ab}\fal{S_0}{q_a}\fal{\gamma}{q_b}\Bigg]
    \nonumber\\
      &=&\frac{G}{4\tilde{V}\chi}\Bigg[
      \left(\Hmbot\right)^2+i\hbar\fal{(\Hmbot-i\hbar A)}{\tau}
      -\frac{i\hbar}{\tilde{V}}\left(\fal{\tilde{V}}{\tau}-(A\tilde{V})
      \right)(\Hmbot-i\hbar A)
    \nonumber\\ &&
      -\frac{2(AW)}{W}(\Hmbot-i\hbar A)
      +\frac{3A\Hmbot}{i\hbar}-2A^2
      \Bigg]\chi\ .
\end{eqnarray}
The definition (\ref{wave functional theta})
of the wave functional $\Theta$ leads to the following relation
for arbitrary derivatives:
\begin{equation}
  \fal{\Theta}{q}=\fal{\chi}{q}\exp\left(\frac{i}{\hbar}\eta G\right)+\mathcal{O}(G)
  =\fal{\chi}{q}\frac{\Theta}{\chi}+\mathcal{O}(G)\ .
\end{equation}
Therefore, the same relation holds for all higher derivatives:
\begin{equation}
  \fal{^n\Theta}{q^n}=\fal{^n\chi}{q^n}\frac{\Theta}{\chi}+\mathcal{O}(G)\ .
\end{equation}
This enables us to rewrite all expressions containing $\chi$ in (\ref{normal part with chi and theta}) in
terms of $\Theta$. We then 
obtain the final result for the normal part of the corrected local Schr\"odinger
equation,
\begin{eqnarray}
\label{final result}
  i\hbar\fal{\Theta}{\tau}&=&\Hmbot\Theta+
  \frac{G}{4\tilde{V}\chi}\Bigg[
      \left(\Hmbot\right)^2+i\hbar\fal{(\Hmbot-i\hbar A)}{\tau}
      -\frac{i\hbar}{\tilde{V}}\left(\fal{\tilde{V}}{\tau}-(A\tilde{V})
      \right)(\Hmbot-i\hbar A)
    \nonumber\\ &&
      -\frac{2(AW)}{W}(\Hmbot-i\hbar A)
      +\frac{3A\Hmbot}{i\hbar}-2A^2
      \Bigg]\Theta\ .
\end{eqnarray}

It would yield a considerable simplification if we had a
vanishing operator $A$.
In particular, the term containing $W$ would be absent.
For a negligible $A$ one would reduce the previous 
expression to
\begin{eqnarray}
\label{final result 2}
  i\hbar\fal{\Theta}{\tau}&=&\Hmbot\Theta
  +\frac{4\pi
  G}{\sqrt{h}\threesR}\Bigg[\left(\Hmbot\right)^2+i\hbar\fal{\Hmbot}{\tau}
  -\frac{i\hbar}{\sqrt{h}\threesR}\fal{(\sqrt{h}\threesR)}{\tau}\Hmbot\Bigg]\Theta\ .
\end{eqnarray}
On a formal level this is  exactly the result that has been obtained from the expansion of the
Wheeler--DeWitt equation. However, there is a difference: The definition of the time functional is different
due to the involvement of the gravitino. But it can be seen that a vanishing gravitino would yield exactly
the same time functional as in the pure bosonic case.
In addition, using the definition (\ref{definition of A}),
this would lead to a vanishing operator $A$. Therefore, the `bosonic limit' of supersymmetric quantum gravity
yields up to the first order of correction terms bosonic canonical quantum gravity. This is a strong argument
for the overall consistency of the supersymmetric theory.

As in \cite{KS91},
the presence of $\Hmbot$ in the above corrections
allows to estimate their importance.
For a Friedmann universe with scale factor $a$ we can
roughly estimate the ratio of the second (and third) to the first correction
term in (\ref{final result 2}),
\begin{equation}
\frac{\hbar}{(\Hmbot)^2}
\frac{\delta \Hmbot}{\delta \tau}
\sim
\frac{\hbar\dot{a}}{(\Hmbot)^2}
\frac{d\Hmbot}{da}
\sim
\frac{\hbar H_0}{E}\ ,
\label{NEW10}
\end{equation}
where $E$ is a typical energy associated with the matter field.
For $E = 700$ GeV and $H_0 = 70$ km/(s\ Mpc) we
obtain approximately $10^{-44}$. Therefore the
quadratic matter Hamiltonian is usually the most
important correction \cite{KS91,BK}. Interesting exceptions
could be very light particles.

Let us add that a violation of unitarity due to the purely imaginary terms
cannot be  immediately concluded. This would require an inner
product that we have not defined here, cf. \cite{OUP}.
Equation (\ref{final result}) is not independent of the chosen
factor ordering.
The explicit representation (\ref{definition of A}) depends on the factor ordering,
since a commutation of $\psi^A_i$ with its derivative changes this term.
Future investigations will deal with the application of
(\ref{final result}) in the context of quantum cosmology and structure
formation.

%%%%%%%%%%%%%%%%%%%%%%%%%%%%%%%%%%%%%%%%%%%%%%%%%%%%%%%%%%%%%%%

\section{Discussion and Outlook}

The purpose of this was paper was to establish a semiclassical approximation scheme for supersymmetric
quantum gravity. 
This has been achieved by extending the Born--Oppenheimer method from the bosonic to the
supersymmetric case. We have considered $N=1$ SUGRA in four spacetime dimensions \cite{SUSY} and performed an
expansion of the Hamiltonian constraint in powers of the gravitational constant by employing its quantum
mechanical operator representation acting on a wave functional of the form (\ref{NEW20}), (\ref{NEW21}). We
have derived, at consecutive orders, the Hamilton--Jacobi equation, the functional Schr\"odinger equation,
and quantum gravitational correction terms to this Schr\"odinger equation.

Within such a framework some relevant features have emerged.
We have obtained explicit formulae to compute the
quantum supersymmetric gravitational corrections that
affect the evolution of the very early universe
during a phase where SUSY plays a crucial role.
This would be of particular relevance for the quantum-to-classical
transition and the ensuing structure formation
\cite{Z14}. We have also found that
 $(i)$ the Hamilton--Jacobi equation and therefore the
background spacetime must involve the gravitino, and
$(ii)$ a (many fingered) local time parameter is
present on $SuperRiem~\Sigma$,
the space of all tetrad and gravitino fields
(plus possible other fields) on a spatial
hypersurface $\Sigma$.

A possible interpretation for that was introduced
and extensively discussed at the end of Section II.
Summarizing it,
 the SUSY Hamilton--Jacobi equation (\ref{susy HJ eq.})  induces
a spacetime background with both tetrad (graviton) and fermionic
(gravitino) terms. It corresponds to a spacetime
metric that will be a solution of the equations of motion
of the theory,
constituting a Grassmann-algebra-valued
field that can be decomposed into the
`body' which takes values in the domain of real or
complex numbers and a `soul' which is
nilpotent \cite{AG,AG1,AG3,AG4}. This
description was introduced
by DeWitt in the context of supermanifold configurations and is
thoroughly described in \cite{AG4}. Hence, a solution of
 the SUSY Hamilton--Jacobi equation (\ref{susy HJ eq.})
 will correspond to a (classical) spacetime, in the
 sense of a  classical
 spacetime with fermionic (gravitino) corrections, 
 leading to a spacetime which can serve as
the appropriate background for the higher orders.

Nevertheless, the proper interpretation of these issues
require more study. A detailed
investigation would, perhaps, require us to follow and extend the
work of Gerlach \cite{Gerlach:69}. More precisely, we should
proceed to consider functionals of the form $\Psi \sim
e^{iS/\hbar}$ and aim to derive the complete set of the equations
of motion of $N=1$ SUGRA in four spacetime dimensions, with
$S$ being a solution of the
 SUSY Hamilton--Jacobi equation (\ref{susy HJ eq.}).
A directly obtained  set of equations should
 be the Hamiltonian equations of motion
 with the presence of Tomonaga's local
 (many fingered) time parameter. Integrating these equations
 on some special hypersurface should
 give the usual SUGRA equations of motion.
The overall procedure
should thus  be checked with respect to the
 limiting case
 without gravitinos (and torsion), that is, with respect to
 general relativity.
 This would provide us with a better understanding 
 of how and what type of spacetime background with
 fermionic corrections emerges, elucidating 
 on the physical meaning of these 
 deviations with respect to the case of canonical 
 general relativity  \cite{OUP,DeWitt:67,Gerlach:69}. 
We intend to address this issue in a future research work.

Somewhat related with the above, there are two additional lines of work to
be considered. In Section IV we have derived the 
quantum gravitational corrections to the Schr\"odinger equation, 
namely normal and tangential correction components. Regarding the former, 
it would be of interest to investigate it further, applying it 
to illustrative minisuperspace case studies, and aiming to determine which 
type of effective quantum field theory and vacuum state are obtained 
as corrections regarding the general relativity case \cite{KS91,Honnef}.
In particular, to analyze if any shift in expectation values of, 
for example, energy levels in a matter Hamiltonian can be produced 
through a SUSY quantum gravitational origin. This would constitute a 
definite prediction from SQC, that is, the SUSY Wheeler--DeWitt equation. 
Even without addressing the issue of regularization, such 
correction terms could lead to quantum gravitational induced shifts, 
observable in principle in the spectrum of the 
cosmic background radiation. Concerning the tangential 
correction component 
(which was not studied in this paper), it would be of interest 
to check if and how it would reflect a breakdown 
of the classical background picture \cite{KS91}, probing 
the superspace environment near a classical solution of 
the SUGRA equations. Moreover, and following the footsteps of 
\cite{BK}, perhaps the use of all constraints, interconnected 
by their constraint algebra, together with these component 
corrections, would allow to generate a Feynman
diagramatic technique involving graviton and gravitino 
loops and vertices, revealing explicitly the back reaction
effects. It could point as well to a correspondence between 
the framework of canonical and covariant SUGRA in a 
semiclassical limit. This is surely a rather 
ambitious line to investigate but we 
think it will provide most elucidating features 
for quantum gravity in general.    

Another pertinent issue to address in the sequence of the 
framework present in this paper is the validity 
of minisuperspace approximation in SQC \cite{Paulo}. 
Different attempts in standard quantum cosmology 
can be found in \cite{KuHu,SiPa}. 
In particular, it was pointed out that the 
minisuperspace approximation in quantum cosmology 
is valid only if the production of gravitons is 
negligible \cite{SiPa}. Hence, it would 
be fairly interesting to establish 
if the presence of fermions 
(gravitinos) and SUSY can either bring additional 
restrictive features on the validity of minisuperspace 
approximation or enlarge the range (through some 
regularization feature) where it can be employed.
We intend to report on this issue in a future publication. 

Finally, the introduction of the Super-DeWitt metric 
in section IV suggests the following possible 
work. In \cite{CKpla}, a connection 
between the sign of the Wheeler--DeWitt metric 
and the attractivity of gravity was studied.
The structure of $SuperRiem~\Sigma$ and its projection down to the
true configurations space was studied for the
bosonic case in \cite{Giulini}.
 It would be of interest to investigate what consequences 
the extra fermion (gravitino) correction terms would
bring into this context.   

%%%%%%%%%%%%%%%%%%%%%%%%%%%%%%%%%%%%%%%%%%%%%%%%%%%%%%%%%%%%%%%%%%%%%%%%
\begin{appendix}
\section{Canonical Quantization of $N=1$ SUGRA}
\label{APP: CQN=1Sugra}

The canonical quantization scheme of
general relativity
starts with the $3+1$ decomposition of spacetime and the reformulation of
the classical action in terms of three-metric and extrinsic curvature. The central role is played by
constraints which reflect the invariances of the classical theory. Upon quantization these constraints lead
to restrictions on the allowed wave functionals \cite{OUP}. In quantum geometrodynamics the central equations
are the quantum Hamiltonian constraint or Wheeler--DeWitt equation and the diffeomorphism (or momentum)
constraints.

In the following we shall summarize the canonical formulation of
$N=1$ SUGRA and its quantization. Details can be found in
\cite{Death,Paulo} and the references therein. 

Dealing with general relativity in
 the presence of fermions requires that one has to work
  with a  tetrad formalism instead of the
metric. This formalism is therefore also needed for SUGRA
where bosons and fermions are treated symmetrically:
for the $N=1$ case we shall have the gravitino as the fermionic
partner to the graviton.
More specifically, at every point of the
spacetime manifold we introduce a pseudo-orthonormal basis $e^\mu_a$ of the tangential space and the
corresponding basis $e^a_\mu$ of the cotangential space, where $a$ is the flat index of the tetrad and runs
from 0 to 3. Indices $a,b,c, \ldots$ are raised and lowered with $\eta^{ab}$ and $\eta_{ab}$, respectively,
where $\eta^{ab}$ has the signature $(-,+,+,+)$. Spacetime indices are raised and lowered with $g^{\mu\nu}$
and $g_{\mu\nu}$, respectively.
 The connection between the spacetime metric
and the internal metric is given by
\begin{equation}
\label{def. of the metric}
  g_{\mu\nu}=\eta_{ab}e^a_{\mu}e^b_\nu
\end{equation}
and
\begin{equation}
  \eta^{ab}=g^{\mu\nu}e^a_{\mu}e^b_{\nu}\ ,
\end{equation}
respectively.

In order to treat the bosonic and fermionic variables as similar as possible, it is appropriate to introduce a
spinorial representation of the tetrad. This is possible since
we can associate in flat space a spinor to any vector
by the Infeld--van der Waerden symbols $\sigma^{AA'}_a$ which are given by
\begin{equation}
  \sigma_0=-\frac{1}{\sqrt{2}}{\mathbb I}\ ,\quad
\sigma_i=\frac{1}{\sqrt{2}}\Sigma_i\ .
\end{equation}
Here, ${\mathbb I}$ denotes the unit matrix
and $\Sigma_i$ are the three Pauli matrices.
The unprimed spinor indices $A,B,C,...$ run from 1 to 2 and the primed spinor indices $A',B',C',...$ take the
values $1'$ and $2'$. The latin indices starting with $i,j,k,...$ assume the values 1, 2, and 3. Hence, the
spinorial version of the tetrad reads
\begin{equation}
  e^{AA'}_\mu=e^a_\mu\sigma^{AA'}_a\ .
\end{equation}
To raise and lower the spinor indices the different representations of the antisymmetric spinorial metric
$\eps^{AB}$, $\eps_{AB}$, $\eps^{A'B'}$ and $\eps_{A'B'}$ are used. Each of them can be written as the same matrix given by
\begin{equation*}
  \left(%
  \begin{array}{cc}
    0 & 1 \\
    -1 & 0 \\
  \end{array}%
\right).
\end{equation*}
Details of this formalism can be found, for example, in \cite{SUSY} and
\cite{SU}.
In curved spacetime, every tensor can be associated with a spinor using the spinorial tetrad,
$
  T^{AA'}=e^{AA'}_\mu T^{\mu},
$
with the inverse relation given by
$
  T^{\mu}=-e_{AA'}^\mu T^{AA'}.
$

For the foliation of spacetime into spatial hypersurfaces we need the future pointing unit normal vector $n^\mu$, whose spinorial version is
\begin{equation}
  n^{AA'}=e^{AA'}_\mu n^{\mu}\ .
\end{equation}
The tetrad is decomposed into the timelike and the spatial components $e^{AA'}_0$ and $e^{AA'}_i$. With the
relation (\ref{def. of the metric}) we find the three-metric
\begin{equation}
  h_{ij}=-e_{{AA'} i}e^{{AA'}}_j =g_{ij}\,.
\end{equation}
This metric and its inverse are used to lower and raise
the spatial indices $i,j,k,\ldots$.
 From the definition of $n^{AA'}$ as a future pointing unit normal to the spatial hypersurfaces $\Sigma$ we obtain the relations
\begin{equation}
  n_{{AA'}}e^{{AA'}}_i=0\quad\textnormal{and}\quad n_{{AA'}}n^{{AA'}}=1\ ,
\end{equation}
which allow to express $n^{{AA'}}$ in terms of $e^{{AA'}}_i$.
An explicit representation is
\begin{equation}
\label{n^A, explicit}
  n^{{AA'}}=\frac{i}{3\sqrt{h}}\eps^{ijk}e^{AB'}_i e_{BB'j}e^{BA'}_k\ ,
\end{equation}
where $h\equiv{\rm det}h_{ij}$. Using the lapse function, $N$,
 and the shift vector, $N^i$,
the timelike component of the tetrad can be decomposed according to
\begin{equation}
  e^{AA'}_0=Nn^{AA'}+N^ie^{AA'}_i\ .
\end{equation}
Further relations are collected in Appendix \ref{APP: Formulas}.

The starting point of the formalism is the action
of $N=1$ SUGRA in four spacetime dimensions \cite{SUSY,vN},
\footnote{If the fields
$e^{AA'}_i$ and $\psi^A_i$ appear in the argument of a functional, the indices are often omitted for simplicity. For example, we
write $S[e,\psi]$ instead of $S[e^{AA'}_i,\psi^A_i]$.}
\begin{equation}
\label{real susy action}
  S[e,\psi]=\int d^4x \left( \frac{1}{16\pi G}{\rm det}(e^a_\mu)R
  +\frac{1}{2}\eps^{\mu\nu\rho\sigma}\left(\bpsi^{A'}_\mu e_{{AA'}\nu}\mathcal{D}_{\rho}\psi_\sigma^A
  +\mathcal{D}_\rho\bar{\psi}^{A'}_\sigma
e_{{{AA'}}\nu}\psi^A_\mu\right)\right)\ ,
\end{equation}
which includes
 the Einstein-Hilbert sector
 (with $\Lambda=0$) and the Rarita-Schwinger component for the gravitino
field $\psi^A_\mu$ with spin $3/2$. The factor ${\rm det}(e^a_\mu)$ equals the square root of the determinant, $\sqrt{-g}$.
The covariant
derivative $\mathcal{D}_\rho$ acts only on the spinor indices and is defined via the spin connection forms
$\omega^{A}_{\hspace{1.6mm} B\rho}$ and $\bar{\omega}^{A'}_{\hspace{1.6mm} B'\rho}$. Their explicit form can
be found in \cite{Death}.
The action (\ref{real susy action}) is invariant under the following local transformations of the
basic fields $e^{AA'}_\mu$ and $\psi^A_\mu$: Local SUSY transformations,
local Lorentz transformations, and local coordinate transformations
(diffeomorphisms).

The canonical fields for the Hamiltonian formulation of
$N=1$ SUGRA  are the spatial components of the
tetrad $e^{{AA'}}_i$ and the gravitino $\psi^{A}_i$ and $\bpsi^{A'}_i$. The momentum conjugate to the tetrad
is defined by
\begin{equation}
  p^{i}_{AA'}=\fal{S}{\dot{e}^{AA'}_{i}}\ ,
\end{equation}
where the dot denotes the partial derivative with respect to the timelike direction. Often a symmetrized
version is used,
\begin{eqnarray}
  \pi^{ij}\equiv-\frac{1}{2}p^{(ij)}\,,\quad p^{ij}=-e^{AA'j}p_{AA'}^i\ .
\end{eqnarray}
The momenta conjugate to the gravitino read
\begin{eqnarray}
\label{momenta conjugate to the gravitino}
    \pi^{i}_A&=&\fal{S}{\dot{\psi}^{A}_{i}}=-\frac{1}{2}\eps^{ijk}\bpsi^{A'}_{b}e_{AA'k}\ ,
  \nonumber\\
    \tilde{\pi}^{i}_{A'}&=&\fal{S}{\dot{\bpsi}^{A'}_{i}}
=\frac{1}{2}\eps^{ijk}\psi^{A}_{b}e_{AA'k}\ .
\end{eqnarray}
We denote the momentum conjugate to $\bpsi^{A'}_i$
by $\tilde{\pi}_{A'}^i$ since it is \emph{minus} the
Hermitian conjugate of $\pi_{A}^i$.
Since no time derivatives occur here, these are constraints which turn out
to be of second class (since their algebra does not close).
We thus have
to formulate Dirac brackets instead of Poisson brackets \cite{Death}.
They read
\begin{eqnarray}
\label{Dirac brackets}
    [e^{AA'}_i(x),e^{BB'}_j(x)]_{*} &=& 0\ ,
  \nonumber\\
    {[e^{AA'}_i(x),p_{BB'}^j (y)]}_{*} &=& \eps^A_{\hspace{1.6mm} B}\eps^{A'}_{\hspace{1.6mm} B'}\delta_i^j\delta(x-y)\ ,
  \nonumber\\
    {[p_{AA'}^{i}{(x)},p_{BB'}^{j}{(y)}]}_{*} &=&
    \frac{1}{4}\big(\eps^{jln}\psi_{Bn}D_{AB'kl}\eps^{ikm}\bpsi_{A'm},
  \nonumber\\&&
    +\eps^{jln}\psi_{Am}D_{BA'lk}\eps^{ikm}\bpsi_{B'n})\delta(x-y)\ ,
  \nonumber\\
    {[\psi^A_i(x),\psi^B_j(y)]}_* &=& 0,
  \nonumber\\
    {[\psi^A_i(x),\bpsi^{A'}_j(y)]}_* &=& -D^{AA'}_{ij}\delta(x-y),
  \nonumber\\
    {[e^{AA'}_i(x),\psi^B_j(y)]}_* &=& 0\ ,
  \nonumber\\
    {[p_{AA'}^{i}(x),\psi^B_j(y)]}_* &=& \frac{1}{2}\eps^{ikl}\psi_{Al}D^{B}_{\hspace{1.6mm} A'jk}\delta(x-y)\ ,
\end{eqnarray}
where
\begin{equation}
\label{A15}
D^{AB'}_{ik}=\frac{-2i}{\sqrt{h}}e^{AC'}_ke_{CC'i}n^{CB'}\ .
\end{equation}
The remaining brackets are obtained by conjugating
the relations containing the field $\psi^A_i$.

Because the action (\ref{real susy action})
is invariant under local Lorentz, SUSY, and coordinate transformations,
the canonical fields are subject to constraints.
Regarding their quantum
representation the following
has to be included. As usual, the classical brackets
(here: the Dirac brackets) are
replaced by $-i/\hbar$ times the commutator or anticommutator of
the corresponding field
operators.
For the Dirac brackets (\ref{Dirac brackets}) this
can be achieved by choosing the following operator
representation of the fundamental fields and
momenta:\footnote{This form for the
representation of the momenta allows the
algebra of the constraints to have a
simpler form; cf. ref. \cite{Death,Paulo} for
further details.}
\begin{eqnarray}
\label{Quantum repr. of the momenta and fields}
    \bpsi^{A'}_i  &=& -i\hbar D^{AA'}_{ji}\fal{}{\psi^A_j}\ ,
  \nonumber\\
    p_{AA'}^{i}  &=&  -i\hbar\fal{}{e^{AA'}_i}-\frac{1}{2}i\hbar\eps^{ijk}\psi_{Aj}D^{B}_{\hspace{1.6mm} A'
    lk}\fal{}{\psi^B_l}\ .
\end{eqnarray}
This is, of course, not the only possible choice.
We can also represent $\psi^A_i$ by a derivative with
respect to $\bpsi^{A'}_i$ if we choose a basis consisting
of eigenstates of $\bpsi^{A'}_i$. But since the
Dirac bracket between $\bpsi^{A'}_i$ and $\bpsi^{A'}_i$
does not vanish, it is not possible to choose a basis
of eigenstates with respect to both of them.

Upon quantization one encounters the usual factor ordering problems.
This is of crucial relevance for the construction of the full theory,
but of less relevance for the present issue of semiclassical approximation.
We shall follow here Ref.~\cite{Death} and do not consider other
possibilities. The quantized Lorentz constraints  read
\begin{eqnarray}
\label{constraint, quantized Lorentz }
  J_{AB} &=& -\frac{i\hbar}{2}\left(e^{A'}_{Ba}\fal{}{e^{AA'}_{a}}+e^{A'}_{Aa}\fal{}{e^{BA'}_a}
            +\psi_{Ba}\fal{}{\psi^{A}}_a+\psi_{Aa}\fal{}{\psi_a^B}\right)\ ,\\
\label{constraint, conj. quantized Lorentz}
  \bar{J}_{A'B'} &=& -\frac{i\hbar}{2}\left(e^{A}_{B'a}\fal{}{e^{AA'}_a}+e^{A}_{A'a}\fal{}{e^{AB'}_a}\right)\ ,
\end{eqnarray}
and the quantized SUSY constraints are given by
\begin{eqnarray}
\label{constraint, quantized susy}
  \bar{S}_{A'} &=& \eps^{ijk}e_{AA'i}\threesD_j\psi^{A}_k+4\pi G\hbar\psi^{A}_i\fal{}{e^{AA'}_i}\ ,\\
\label{constraint, conj. quantized susy}
  S_A &=& i\hbar\threesD_i\left(\fal{}{\psi^A_i}\right)+4\pi iG\hbar\fal{}{e^{AA'}_i}\left(D_{ji}^{BA'}\fal{}{\psi^B_j}\right)\ .
\end{eqnarray}
Calculating the anticommutator between the SUSY constraints yields
\begin{equation}
\label{commutator of S_A, S_A'}
  [S_A(x),\bar{S}_{A'}(y)]_+ =4\pi G\hbar\mathcal{H}_{AA'}(x)\delta(x,y)\ ,
\end{equation}
with
\begin{eqnarray}
  \label{constraint, quantized HAA, rept}
    \mathcal{H}_{AA'} &=&  4\pi G i
    \hbar^2\psi^B_i\fal{}{e^{AB'}_j}\left[\eps^{ilm}D_{B\hspace{1.5mm}mj}^{\hspace{1.5mm}B'}D^C_{\hspace{1.5mm}A'kl}
    \fal{}{\psi^C_k}\right]
  \nonumber\\ &&
    -4\pi G i\hbar^2\fal{}{e^{AB'}_j}\left[\DBij\fal{}{e^{BA'}_i}\right]
  \nonumber\\ &&
    - \frac{i\hbar}{2}\eps^{ijk}\left[\left({\threesD}_j\psi_{Ak}\right)\DBAli\fal{}{\psi^B_l}+
    \psi_{Ai}\left({\threesD}_j\DBAlk\fal{}{\psi^B_l}\right)\right]
  \nonumber\\ &&
    - {i\hbar\,\threesD}_i\left(\fal{}{e^{AA'}_i}
    +\frac{1}{2}\eps^{ijk}\psi_{Aj}D^{B}_{\hspace{1.5mm}A'lk}\fal{}{\psi^B_l}\right)
    +n_{AA'}\frac{1}{G}V[e]\ ,
\end{eqnarray}
where $V[e]=\sqrt{h}\threesR/16\pi$.
Note that from $\threeD_j$ (denoting a spatial covariant derivative
acting on the spinor indices),
\begin{equation}
 \threeD_j T^{AA'} = \partial_jT^{AA'}
+\threeom^{A}_{B}T^{BA'}+\threebarom^{A'}_{B'}T^{AB'}\ ,
\end{equation}
where $\threeom^{A}_{B}$ and $\threebarom^{A'}_{B'}$ are the two parts of
the spin connection, see (\ref{APP: spin
conn., decomp. in primed and unprimed}), we obtain,
by decomposing the three-dimensional spin connection
$\threeom^{AA'BB'}_i$ contained in the covariant derivative $\threeD_j$ into
a pure bosonic part and the
contorsion (\ref{APP: spin conn., decomposition in contorsion and torsion-free}),
\begin{eqnarray}
 \threeom^{AA'BB'}_i=\threesom^{AA'BB'}_i+\threekappa^{AA'BB'}_i.
\end{eqnarray}
The torsion-free derivative is denoted by $\threesD_j$.
This also leads to  simpler versions of the SUSY
constraints \cite{Death},
where
$\bar{S}_{A'}$ is the Hermitian conjugate of
$S_A$.
They guarantee the
invariance of the action under left- and right-handed SUSY
transformations, respectively. Note that no torsion terms appear there.
Moreover,
$\threeR$ is the three-dimensional scalar curvature
(\ref{APP: scalar curv.}).

The calculation leading to (\ref{commutator of S_A, S_A'}) shows that this
factor ordering does not lead to quantum anomalies, at least not on
a formal level. The expression of the right-hand side of
(\ref{commutator of S_A, S_A'}) can
be interpreted
as  a combination of the  Hamiltonian
and momentum constraints obtained
from the action of $N=1$ SUGRA through
variational methods plus combinations of the Lorentz constraints.
A solution of the above quantum SUSY constraints must thus automatically obey
the other constraints. It is an unsolved issue whether the full
quantum algebra of constraints is free of anomalies. Calculations
in \cite{Wulf} seem to indicate that anomalies may occur in the
commutators of the SUSY constraints with $\mathcal{H}_{AA'}(x)$.
A definite statement can, however, only be made after a rigorous
regularization scheme has been employed.
We
assume in this paper that anomalies are absent. The question of anomalies is an open issue
in all approaches of canonical quantum gravity \cite{NPZ}.

%%%%%%%%%%%%%%%%%%%%%%%%%%%%%%%%%%%%%%%%%%%%%%%%%%%%%%%%%%%%%%%%%%%%%%%%%%

\section{The semiclassical approximation scheme
for canonical quantum gravity}
\label{Scheme}

This appendix contains a brief review of the semiclassical approximation
scheme, as it has been applied on a formal level to quantum
geometrodynamics \cite{OUP}. This will enable us in particular to
make a comparison with the SUSY case discussed herein this article.

Our starting point is the full Wheeler--DeWitt equation
and the momentum constraints,
\begin{eqnarray}
  \label{Wheeler-DeWitt eq.}
  \left(-16\pi G\hbar^2G_{ijkl}\zfal{}{h_{ij}}{h_{kl}}-\frac{1}{16\pi G}\sqrt{h}\threeR
  +\Hmbot\right)\Psi[h_{ij},\Phi]&=&0\,,
  \\
  \label{momentum constraints}
  \left(-\frac{2i}{\hbar}\threenabla_j h_{ik}\fal{}{h_{jk}}+\mathcal{H}^{\rm{m}}_i\right)\Psi[h_{ij},\Phi]&=&0\,,
\end{eqnarray}
where $\Phi$ denotes here a general non-gravitational field.
It is convenient to introduce the parameter
\begin{displaymath}
  M\equiv\frac{1}{32{\pi}G}
\end{displaymath}
and perform an expansion with respect to $M$. Although $M$ does not
have the dimension of a mass (it is proportional to the Planck mass
squared),
it brings the Wheeler--DeWitt equation into
a form similar to the Schr\"odinger equation in quantum mechanics and
thus allows the (formal) application of the Born--Oppenheimer scheme
\cite{OUP,Honnef}.\footnote{In quantum electrodynamics, one can perform
an expansion with respect to the electric charge \cite{KPS}.}
More generally, the approximation scheme starts with a division into
`slow' and `fast' degrees of freedom. An expansion with respect to $M$
is the simplest way to implement this idea, in that the gravitational
variables are `slow' and the remaining (`matter') variables (whose
Hamiltonian is denoted by $\Hmbot$) are `fast'.
Equation (\ref{Wheeler-DeWitt eq.}) then becomes
\begin{equation}
\label{Wheeler-DeWitt eq. with M}
  \left(-\frac{\hbar^2}{2M}G_{ijkl}\zfal{}{h_{ij}}{h_{jk}}+MV^{\rm g}
  +\Hmbot\right)\Psi=0\ ,
\end{equation}
with $V^{\rm g}=-2\sqrt{h}\threeR$.
For the matter Hamiltonian density $\Hmbot$ we assume for simplicity a
minimally coupled scalar field $\Phi$.

Making for the wave functional the ansatz,
\begin{equation}
  \Psi[h_{ij},\Phi]=\exp\left(\frac{i}{\hbar}S[h_{ij},\Phi]\right)
\end{equation}
and expanding
\begin{displaymath}
  S[h_{ij},\Phi]=\sum_{n=0}^{\infty}S_n[h_{ij},\Phi]M^{-n+1}\ ,
\end{displaymath}
we find from (\ref{Wheeler-DeWitt eq. with M})
several relevant equations at consecutive orders
of $M$.

The highest order ($M^2$) expresses the independence of
$S_0$ on the matter field $\Phi$, that is, $S_0\equiv{S_0[h_{ij}]}$.
The next order ($M^1$) yields the
Hamilton--Jacobi equation for the gravitational field,
\begin{equation}
  \label{bos. HJ eq.}
  \frac{1}{2}G_{ijkl}\fal{S_0}{h_{ij}}\fal{S_0}{h_{kl}}+V^{\rm g}=0\ .
\end{equation}
Actually, (\ref{bos. HJ eq.}) represents an infinite number of equations, one at every point of space. In
addition we have to expand the momentum constraints (\ref{momentum constraints}) and obtain
\begin{equation}
\label{exp. mom.constraints}
  h_{ij}\threenabla_k\left(\fal{S_0}{h_{ik}}\right)=0\ .
\end{equation}
Every solution of (\ref{bos. HJ eq.}) determines a family of solutions of the classical field equations. Equations
(\ref{bos. HJ eq.}) and (\ref{exp. mom.constraints})
are equivalent to Einstein's field equations \cite{DeWitt:67,Gerlach:69}.

The next order ($M^0$) can be simplified by defining the wave functional
\begin{equation}
  \label{wave functional chi}
  \chi=D[h_{ij}]\exp\left(\frac{iS_1[h_{ij},\Phi]}{\hbar}\right)\ .
\end{equation}
Choosing for the `WKB prefactor' $D$ the `conservation law'
(which in quantum mechanics would just express the conservation of
probability)
\begin{equation}
  \label{conservation law}
  G_{ijkl}\fal{}{h_{ij}}\left(\frac{1}{D^2}\fal{S_0}{h_{kl}}\right)=0\ ,
\end{equation}
the equation at this order becomes the `Tomonaga--Schwinger equation'
or `local Schr\"odinger equation'
\begin{equation}
\label{functional Schroedinger eq.}
  i\hbar G_{ijkl}\fal{S_0}{h_{ij}}\fal{\chi}{h_{kl}}\equiv i\hbar\fal{\chi}{\tau}=\Hmbot\chi\ ,
\end{equation}
where the time functional $\tau$ is implicitly defined by
\begin{equation}
\label{Time functional}
  G_{ijkl}(x)\fal{S_0}{h_{ij}(x)}\fal{\tau(y;h_{ij}]}{h_{kl}(x)}=\delta(x-y)\ .
\end{equation}
`Time' is thus defined through the chosen solution $S_0$ of
the Hamilton--Jacobi equation. In fact, $\tau$ is not a spacetime scalar,
but the semiclassical scheme can
nevertheless be consistently defined \cite{GK}.
The (functional) Schr\"odinger equation is found upon integrating
(\ref{Time functional}) over three dimensional space.

The next order ($M^{-1}$) yields quantum gravitational correction terms
to (\ref{functional Schroedinger eq.}). In \cite{KS91} only those
correction terms were considered that act along the chosen
classical spacetime; those terms appear to be the dominating one.
In \cite{BK} all correction terms were treated in
great detail. In the present case we  followed the treatment
in \cite{KS91} in order to show the essential features of the
semiclassical approximation scheme.

%%%%%%%%%%%%%%%%%%%%%%%%%%%%%%%%%%%%%%%%%%%%%%%%%%%%%%%%%%%%%%%%%%%%%%%

\section{Formulae used for the calculation of supersymmetric expressions}
\label{APP: Formulas}
\subsection{General formulae}
In Appendix A, we have chosen the signature of the four-metric
$g_{\mu\nu}$ as $(-,+,+,+)$. Therefore, the metric $h_{ij}$ on the spacelike hypersurfaces has the signature
$(+,+,+)$ which gives a positive determinant. Thus the three-dimensional total antisymmetric tensor density
can be defined by $\eps^{123}=\eps_{123}=+1$. Using this definition,
we have for the timelike normal vector $n_{AA'}$
and the tetrad $e^{AA'}_i$ the relations
\begin{eqnarray}
  \label{APP: SUSY-relation 1}
    n_{AA'}n^{AB'} &=&\frac{1}{2}\eps_{A'}^{\hspace{1.6mm} B'},
  \\
  \label{APP: SUSY-relation 2}
    n_{AA'}n^{BA'} &=& \frac{1}{2}\eps_{A}^{\hspace{1.6mm} B},
  \\
  \label{APP: SUSY-relation 3}
    e_{AA'i}e^{AB'}_j &=& -\frac{1}{2}h_{ij}\eps_{A'}^{\hspace{1.6mm} B'}-i\sqrt{h}\eps_{ijk}n_{AA'}e^{AB'k},
  \\
  \label{APP: SUSY-relation 4}
    e_{AA'i}e^{BA'}_j &=& -\frac{1}{2}h_{ij}\eps_{A}^{\hspace{1.6mm}
    B}-i\frac{1}{\sqrt{h}}\eps_{ijk}n_{AA'}e^{BA'k},
  \\
  \label{APP: SUSY-relation 5}
    e_{AA' i}e_{BB'}^i &=& n_{AA'}n_{BB'}-\eps_{AB}\eps_{A'B'}\ .
\end{eqnarray}
 From equations (\ref{APP: SUSY-relation 3}) and (\ref{APP: SUSY-relation 4}) we obtain by contracting with
$\eps^{ijl}$,
\begin{eqnarray}
    n_{AA'}e^{AB'l} & = & -n^{AB'}e_{AA'}^k  = \frac{i}{2\sqrt{h}}\eps^{ijl}e_{AA'i}e^{AB'}_j,\\
    n_{AA'}e^{BA'l} & = & -n^{BA'}e_{AA'}^k  = -\frac{i}{2\sqrt{h}}\eps^{ijl}e_{AA'i}e^{BA'}_j\ .
\end{eqnarray}

The three-dimensional torsion-free spin connection $\threesom^{{AA'}{BB'}}_i$ can be expressed in terms of
$n^{{AA'}}$ and $e^{{AA'}}_{i}$ \cite{Death},
\begin{eqnarray}
\label{APP: spin conn., torsion free explicit}
   & &  \threesom^{{AA'}{BB'}}_{i}= e^{{BB'} j}\partial_{[j}e^{{AA'}}_{i]}
\nonumber\\ & &
    -\frac{1}{2}\left(e^{{AA'} j}e^{{BB'} k}e^{{CC'}}_i\partial_je_{{CC'} k}
    +e^{{AA'} j}n^{BB'} n^{CC'} \partial_j e_{{CC'} i}+n^{AA'}\partial_in^{BB'}\right)
  \nonumber\\ \; &&
  \;  -e^{{AA'} j}\partial_{[j}e^{{BB'}}_{i]}\nonumber \\ & &
+\frac{1}{2}\left(e^{{BB'} j}e^{{BB'} k}e^{{CC'}}_i\partial_je_{{CC'} k}
    +e^{{BB'} j}n^{AA'} n^{CC'} \partial_j e_{{CC'} i}+n^{BB'}\partial_in^{AA'}\right)\ .\hspace{4mm}
\end{eqnarray}
The four-dimensional torsion is given by
\begin{equation}
  S^{AA'}_{\mu\nu}=-4\pi i G\bpsi^{A'}_{[\mu}\psi^{A}_{\nu]}\ ,
\end{equation}
and its tensorial version reads
\begin{equation}
  S^{\rho}_{\hspace{1.6mm}\mu\nu}=-e_{AA'}^{\rho}S^{AA'}_{\mu\nu}.
\end{equation}
The contorsion tensor $\kappa$ is defined by
\begin{equation}
  \kappa_{\mu\nu\rho}=S_{\nu\mu\rho}+S_{\rho\nu\mu}+S_{\mu\nu\rho}\ .
\end{equation}
The three-dimensional contorsion is simply obtained by restriction of the four-dimensional quantity,
\begin{equation}
  \threekappa_{ijk}=\kappa_{ijk}\ .
\end{equation}
With the spinorial contorsion $\threekappa^{{AA'}{BB'} i}=e^{{AA'}
j}e^{{BB'}}_k\threekappa_{jki}=-\threekappa^{{BB'}{AA'} i}$, the spin connection reads
\begin{equation}
\label{APP: spin conn., decomposition in contorsion and torsion-free}
  \threeom^{{AA'}{BB'}}_i=\threesom^{{AA'}{BB'}}_i+\threekappa^{{AA'}{BB'}}_i.
\end{equation}
It can be decomposed into a primed and an unprimed part:
\begin{equation}
  \threeom^{AA'BB'}_i=\threeom^{AB}_i\bar{\eps}^{A'B'}+\threebarom^{A'B'}_i\eps^{AB}.
\end{equation}
Using the antisymmetry $\threeom^{AA'BB'}_i=-\threeom^{BB'AA'}_i$, we obtain the symmetries
$\threeom^{AB}_i=\threeom^{BA}_i$ and $\threebarom^{A'B'}_i=\threebarom^{B'A'}_i$ and the explicit
representations
\begin{equation}
\label{APP: spin conn., decomp. in primed and unprimed}
  \threeom^{AB}_i=\frac{1}{2}\threeom^{A\hspace{1.6mm} BB'}_{\hspace{1.6mm} B'i},\quad
  \threebarom^{A'B'}_i=\frac{1}{2}\threeom^{\hspace{1.6mm} A'BB'}_{Bi}.
\end{equation}
Analogous relations hold for $\threesom^{{AA'}{BB'}}_i$ and $\threekappa^{{AA'}{BB'}}_i$. The components of
the three-dimensional curvature in terms of the spin connection read
\begin{eqnarray}
    \threeR^{AB}_{ij}&=&2\left(\partial_{[i}\threeom^{AB}_{j]}+\threeom^{A}_{C[i}\threeom^{CB}_{j]}\right)\ ,
  \nonumber\\
    \threebarR^{A'B'}_{ij}&=&2\left(\partial_{[i}\threebarom^{A'B'}_{j]}
    +\threebarom^{A'}_{C'[i}\threebarom^{C'B'}_{j]}\right).
\end{eqnarray}
Because of the symmetry of $\threeom^{[AB]}_i=0$ and $\threebarom^{[A'B']}_i=0$, the chosen notation
$\threeom^{A}_{B i}$ and $\threebarom^{A'}_{B'i}$ is unambiguous. The horizontal position of the indices does
not need to be fixed. The scalar curvature is given by
\begin{equation}
\label{APP: scalar curv.}
  \threeR=e_{AA'}^ie_{BB'}^j(\threeR^{AB}_{ij}\bar{\eps}^{A'B'}+\threebarR^{A'B'}\eps^{AB})\ .
\end{equation}
The same procedure performed on $\threesom^{{AA'}{BB'}}_i$ leads to the torsion-free scalar curvature,
\begin{eqnarray}
    \threesR^{AB}_{ij}&=&2\left(\partial_{[i}\threesom^{AB}_{j]}+\threesom^{A}_{C[i}\threesom^{CB}_{j]}\right),
  \nonumber\\
    \threesbarR^{A'B'}_{ij}&=&2\left(\partial_{[i}\threesbarom^{A'B'}_{j]}
    +\threesbarom^{A'}_{C'[i}\threesbarom^{C'B'}_{j]}\right),
\end{eqnarray}
and
\begin{equation}
\label{APP: scalar curv., torsion free}
  \threesR=e_{AA'}^ie_{BB'}^j(\threesR^{AB}_{ij}\eps^{A'B'}+\threesbarR^{A'B'}\eps^{AB})\ .
\end{equation}

\subsection{Equations used in Section \ref{HamiltonJacobi}} In Section \ref{HamiltonJacobi} we need the
explicit form of the expressions
\begin{equation}
  \eps^{ilm}n^{AA'}\fal{}{e^{AB'}_j}(\DD)
\end{equation}
and
\begin{equation}
  n^{{AA'}}\fal{}{e^{AB'}_j}\DBij.
\end{equation}
To evaluate these terms we first need an explicit form of $\delta{n^{{AA'}}}/\delta{e^{{BB'}}_j}$. Of course,
for this purpose relation (\ref{n^A, explicit}), which expresses $n^{AA'}$ in terms of the tetrad, can be
used, but it is more convenient to start from $n^{{AA'}}e_{{{AA'}}i}=0$:
\begin{equation}
  0=e^{{{CC'}}i}\fal{n^{{AA'}}e_{{AA'}}i}{e^{{BB'}}_j}=
  n^{{CC'}}n_{{AA'}}\fal{n^{{AA'}}}{e^{{BB'}}_j} -\eps_A^{\hspace{1.6mm} C}\eps_{A'}^{\hspace{1.6mm}
  C'}\fal{n^{{AA'}}}{e^{{BB'}}_j}+e^{{{CC'}}j}n_{{BB'}}\ .
\end{equation}
In addition we have
\begin{equation}
  \fal{n^{{AA'}}}{e^{{BB'}}_j}=\fal{n^{{CC'}}n_{{CC'}}n^{{AA'}}}{e^{{BB'}}_j}
  =2n_{{CC'}}n^{{AA'}}\fal{n^{{AA'}}}{e^{{BB'}}_j}+\fal{n^{{AA'}}}{e^{{BB'}}_j}
\end{equation}
and obtain
\begin{equation}
  \fal{n^{{AA'}}}{e^{{BB'}}_j}=e^{{AA'} j}n_{{BB'}}\ .
\end{equation}
We often need the derivative of the determinant $h$ of the three-metric,
\begin{equation}
  \frac{\partial h}{\partial h_{ij}}=h^{ij}h\ .
\end{equation}
Therefore we get
\begin{equation}
  \fal{h}{e^{AA'}_i}=-2he_{AA'}^i\ .
\end{equation}

Using this as well as (\ref{APP: SUSY-relation 1})--(\ref{APP: SUSY-relation 5}), we are able to calculate
 expressions (\ref{derivative of epsDD}) and (\ref{derivative of DBij}):
\begin{eqnarray}
    n^{AA'}\eps^{ilm}\fal{}{e^{AB'}_j}(\DD) &=&
    -4n^{AA'}\eps^{ilm}\fal{}{e^{AB'}_j}
    \left(\frac{1}{h}e_{Bj}^{\hspace{1.6mm} D'}e_{DD'm}n^{DB'}e^{CE'}_le_{EE'k}n^{E}_{\hspace{1.6mm} A'}\right)
  \nonumber\\
    &\hspace{-6cm}=&\hspace{-3cm}
    \eps_B^{\hspace{1.6mm} C}\delta^i_k\frac{i}{\sqrt{h}}\left(1-1+\frac{1}{2}-\frac{1}{2}\right)
    +\frac{2i}{\sqrt{h}}\left(2e^{CB'i}e_{BB'k}+e_{BB'}^ie^{CB'}_k\right)
  \nonumber\\
    &\hspace{-6cm}=&\hspace{-3cm}
    \frac{-3i}{\sqrt{h}}\delta^i_k\eps_B^{\hspace{1.6mm} C}-2h^{ij}\eps_{jkl}n^{CB'}e_{BB'}^l\ ,
\end{eqnarray}
\begin{eqnarray}
    n^{AA'}\fal{}{e^{AB'}_j}\DBij &=& -2in^{AA'}\fal{}{e^{AB'}_j}
    \left(\frac{1}{\sqrt{h}}e^{BC'}_je_{CC'i}n^{CB'}\right)
  \nonumber\\
    &=&-\frac{2i}{\sqrt{h}}n^{AA'}n^{BC'}e_{AC'i}\ .
\end{eqnarray}

In order to compare the results in Sections \ref{HamiltonJacobi}, \ref{Schroedinger}, and \ref{Corrections}
with those in Appendix \ref{Scheme}, we need some rules for the transformation of formulae in terms of the
tetrad ${e^{AA'}_i}$ into formulae in terms of the three-metric $h_{ij}$. Let $\mathcal{F}[e]$ be a functional
depending on the tetrad. Indeed, $h_{ij}$ can be expressed in terms of the tetrad, since we have the relation
$h_{ij}=-e^{AA'}_ie_{{AA'}_j}$, but an inverse relation does of course
not exist. We have therefore to restrict the
functional $\mathcal{F}$. We must demand that it can be written in the form
$\mathcal{F}[h_{ij}]$. Then we find for the transformation of the functional derivatives, by using the chain
rule,
\begin{eqnarray}
\label{APP: trafo from e to h}
  \fal{\mathcal{F}}{e^{AA'}_i}
  & = & \fal{\mathcal{F}}{h_{jk}}\fal{h_{jk}}{e^{AA'}_i}
  = -\fal{\mathcal{F}}{h_{jk}}\eps_{BC}\eps_{B'C'}\fal{e^{BB'}_{j} e^{CC'}_{k}}{e^{AA'}_i} \nonumber\\
  & = & -\fal{\mathcal{F}}{h_{ik}}\eps_{AC}\eps_{A'C'}e^{CC'}_{k}-\fal{\mathcal{F}}{h_{ji}}\eps_{BA}\eps_{B'A'}e^{BB'}_{j}
  = -2\fal{\mathcal{F}}{h_{ij}}e_{{AA'}{j}}\ .
\end{eqnarray}
Using $e^{{AA'} i}e_{{AA'} j}=-\delta^{i}_j$, the inverse relation can be read off immediately. It holds for
an arbitrary functional $\mathcal{G}[h_{ij}]$ without any restrictions, since it is always possible to rewrite
$\mathcal{G}[{h_{ij}}]$ in the form $\mathcal{G}[e]$,
\begin{equation}
\label{APP: trafo from h to e}
  \fal{\mathcal{G}}{h_{ij}}=\frac{1}{2}e^{{AA'} j}
\fal{\mathcal{G}}{e^{AA'}_i}\ .
\end{equation}

\section{Relations used for the corrections of the Schr\"odinger equation}

In Section \ref{Corrections} we have calculated the corrections 
of the Schr\"odinger equation at order $G^1$.
To obtain the explicit form of the correction terms, the following relations are used. For the treatment of
terms (i), (ii), (iii), and (iv) in (\ref{contributions in G^1}) we need
\begin{eqnarray}
\label{A1}
    \frac{1}{\chi}\zfal{\chi}{{e^{AB'}_j}}{\psi^C_k}&=&
    \frac{1}{W}\zfal{W}{{e^{AB'}_j}}{\psi^C_k}+\frac{i}{\hbar W}\fal{W}{{e^{AB'}_j}}\fal{S_1}{\psi^C_k}
    +\frac{i}{\hbar W}\fal{S_1}{{e^{AB'}_j}}\fal{W}{\psi^C_k}
  \nonumber\\ &&
    +\frac{i}{\hbar}\zfal{S_1}{{e^{AB'}_j}}{\psi^C_k}-\frac{1}{\hbar^2}\fal{S_1}{{e^{AB'}_j}}\fal{S_1}{\psi^C_k}\ ,
\end{eqnarray}
\begin{eqnarray}
\label{A2}
    \frac{1}{\chi}\fal{\chi}{\psi^C_k}
    = \frac{1}{W}\fal{W}{\psi^C_k}+\frac{i}{\hbar}\fal{S_1}{\psi^C_k}\ ,
\end{eqnarray}
\begin{eqnarray}
\label{A3}
  \frac{1}{\chi}\fal{W}{{e^{AB'}_j}}\fal{\chi}{\psi^C_k}
  = \frac{1}{W}\fal{W}{{e^{AB'}_j}}\fal{W}{\psi^C_k} + \frac{i}{\hbar}\fal{W}{{e^{AB'}_j}}\fal{S_1}{\psi^C_k}\ ,
\end{eqnarray}
and
\begin{eqnarray}
\label{A4}
  \frac{1}{\chi}\fal{\chi}{{e^{AB'}_j}}\fal{W}{\psi^C_k}
  = \frac{1}{W}\fal{W}{{e^{AB'}_j}}\fal{W}{\psi^C_k}+\frac{i}{\hbar}\fal{S_1}{{e^{AB'}_j}}\fal{W}{\psi^C_k}\ .
\end{eqnarray}
For parts (v)--(viii) of (\ref{contribution in G^1, second version}), we use
\begin{eqnarray}
\label{A5}
    \frac{1}{\chi}\zfal{\chi}{{e^{AB'}_j}}{{e^{BA'}_i}}&=&
    \frac{1}{W}\zfal{W}{{e^{AB'}_j}}{{e^{BA'}_i}}+\frac{i}{\hbar W}\fal{W}{{e^{AB'}_j}}\fal{S_1}{{e^{BA'}_i}}
    +\frac{i}{\hbar W}\fal{S_1}{{e^{AB'}_j}}\fal{W}{{e^{BA'}_i}}
  \nonumber\\ &&
    +\frac{i}{\hbar}\zfal{S_1}{{e^{AB'}_j}}{{e^{BA'}_i}}-\frac{1}{\hbar}\fal{S_1}{{e^{AB'}_j}}\fal{S_1}{{e^{BA'}_i}}
\end{eqnarray}
and
\begin{eqnarray}
\label{A6}
  \frac{1}{\chi}\fal{\chi}{{e^{BA'}_i}}=
  \frac{1}{W}\fal{W}{{e^{BA'}_i}}+\frac{i}{\hbar}{{e^{AB'}_j}}\fal{S_1}{{e^{BA'}_i}}\ ,
\end{eqnarray}
\begin{eqnarray}
\label{A7}
  \frac{1}{\chi}\fal{W}{{e^{AB'}_j}}\fal{\chi}{{e^{BA'}_i}}=
  \frac{1}{W}\fal{W}{{e^{AB'}_j}}\fal{W}{{e^{BA'}_i}}
  +\frac{i}{\hbar}\fal{W}{{e^{AB'}_j}}\fal{S_1}{{e^{BA'}_i}}\ ,
\end{eqnarray}
\begin{eqnarray}
\label{A8}
  \frac{1}{\chi}\fal{\chi}{{e^{AB'}_j}}\fal{W}{{e^{BA'}_i}}=
  \frac{1}{W}\fal{W}{{e^{AB'}_j}}\fal{W}{{e^{BA'}_i}}
  +\frac{i}{\hbar}\fal{S_1}{{e^{AB'}_j}}\fal{W}{{e^{BA'}_i}}\ .
\end{eqnarray}
To perform the next step we need
\begin{eqnarray}
\label{A9}
  i\hbar\fal{\Theta}{\tau}=i\hbar\exp\left(\frac{i}{\hbar}\eta G\right)\fal{\chi}{\tau}
  -G\chi\exp\left(\frac{i}{\hbar}\eta G\right)\fal{\eta}{\tau}\ .
\end{eqnarray}
The last term in (\ref{contribution in G^1, second version}) 
is part of the expression
\begin{eqnarray}
\label{A10}
  \Hmbot\Theta =\exp\left(\frac{i}{\hbar}\eta G\right)\Hmbot\chi
  -\frac{i\hbar G}{2\sqrt{h}}\left(\frac{2}{\chi} \fal{\chi}{\Phi}\fal{\eta}{\Phi}
  +\fal{^2\eta}{\Phi^2}\right)\Theta
  +\mathcal{O}(G^2)\ .
\end{eqnarray}
Since we are only interested in corrections of order $G$, we neglect terms of order $G^2$.
\end{appendix}

%%%%%%%%%%%%%%%%%%%%%%%%%%%%%%%%%%%%%%%%%%%%%%%%%%%%%%%%%%%%%%%%%%%%%%%%%%%%%%%%%%%%%%%%%%%%%%%%

\section*{Acknowledgements}

This work was supported by DAAD-GRICES/2004/2005 - D/03/40416,
CRUP-AI-A-21/2004 as
well as the grants POCTI(FEDER) P - FIS - 57547/2004 and
  CERN P - FIS - 49529/2003. P.V.M. is supported by the grant
   FCT (FEDER) SFRH - BSAB 396/2003. He also wishes to
    thank Queen Mary College, London,
  for kind hospitality  during his sabbatical and
  expresses his warmest thanks to the Institut f\"ur Theoretische Physik,
  Universit\"at zu K\"oln, for kind hospitality   during his
  visits. C.K. and T.L. acknowledge kind hospitality at the Universidade
de Beira Interior, Covilh\~a.

%%%%%%%%%%%%%%%%%%%%%%%%%%%%%%%%%%%%%%%%%%%%%%%%%%%%%%%%%%%%%%%%%%%%%


\begin{thebibliography}{99}
\bibitem{OUP} C. Kiefer, {\em Quantum Gravity} (Clarendon Press, Oxford,
 2004).

\bibitem{CQg} D. Giulini, C. Kiefer, and C. L\"ammerzahl (Eds.),
 {\em Quantum Gravity}. Lecture Notes in Physics~631 (Springer, Berlin,
 2003); C. Rovelli, {\em Quantum Gravity} (Cambridge University Press,
  Cambridge, 2004).

\bibitem{SSt}  J. Polchinski, {\em String Theory},
 two volumes (Cambridge University Press, 1998).

\bibitem{SUSY} S. Weinberg, {\em The quantum theory of fields,
Vol. III (Supersymmetry)} (Cambridge University Press, Cambridge, 2000);
J. Wess and J. Bagger, \textit{Supersymmetry and
Supergravity}, second edition (Princeton University Press, 1992).

\bibitem{Death} P. D. D'Eath, {\em Supersymmetric quantum cosmology}
(Cambridge University Press, Cambridge, 1996);
P. D. D'Eath, Phys. Rev. D {\bf 29}, 2199 (1984).

\bibitem{Paulo}P. V. Moniz, Int. J. Mod. Phys. A {\bf 11}, 4321 (1996).

\bibitem{Moniz:1996qz}
  P.~V.~Moniz,
   Nucl.\ Phys.\ Proc.\ Suppl.\  {\bf 57}, 307 (1997)
  and references therein.
  
  \bibitem{SQC1} A.~Macias, O.~Obregon and M.~P.~Ryan,
   Class.\ Quant.\ Grav.\  {\bf 4} 1477 (1987); 
   R.~Graham,
    Phys.\ Rev.\ Lett.\  {\bf 67}, 1381 (1991); 
 J.~Bene and R.~Graham,
  Phys.\ Rev.\ D {\bf 49}, 799 (1994);
    P. V. Moniz, Gen. Rel. Grav. \textbf{28} 97, (1996); 
P. V. Moniz, A Supersymmetric {\em Vista} for 
Quantum Cosmology, to appear in Gen. Rel. Grav. 
(2005), issue dedicated to M. Ryan's 65th birthday, edited by
C. L\"ammerzahl, A. Macias, and O. Obregon.  

\bibitem{SQC2} A. D. Y. Cheng and P. V. Moniz, Int. J. Mod. Phys.
\textbf{D4}, 189 (1995); P. V. Moniz, 
Int. J. Mod. Phys. A \textbf{11}, 1763 (1996);
 Int. J. Mod. Phys. D \textbf{6}, 465 (1997);
R.~Graham and H.~Luckock,
   Phys.\ Rev.\ D {\bf 49}, R4981 (1994);
 R.~Graham and J.~Bene,
  Phys.\ Lett.\ B {\bf 302}, 183 (1993);
A.~Csordas and R.~Graham,
   Phys.\ Rev.\ Lett.\  {\bf 74}, 4129 (1995);
     J.~Socorro, O.~Obregon, and A.~Macias,
   Phys.\ Rev.\ D {\bf 45}, 2026 (1992);
   P.~D.~D'Eath, S.~W.~Hawking, and O.~Obregon,
  Phys.\ Lett.\ B {\bf 300} 44 (1993);
 P.~D.~D'Eath and D.~I.~Hughes,
   Nucl.\ Phys.\ B {\bf 378}, 381 (1992);
P.~D.~D'Eath,
   Phys.\ Rev.\ D {\bf 48}, 713 (1993);
J.~Socorro and E.~R.~Medina,
   Phys.\ Rev.\ D {\bf 61}, 087702 (2000);
  J.~Socorro,
   Rev.\ Mex.\ Fis.\  {\bf 48}, 112 (2002).
  
  \bibitem{SQC3} J.~E.~Lidsey,
   Phys.\ Rev.\ D {\bf 52}, R5407 (1995);
J.~E.~Lidsey and P.~V.~Moniz,
   Class.\ Quant.\ Grav.\  {\bf 17}, 4823 (2000);
  P.~V.~Moniz,
   Nucl.\ Phys.\ Proc.\ Suppl.\  {\bf 88}, 57 (2000).
  
  \bibitem{SQC4a} P.~V.~Moniz,
  Ann. Phys. (Leipzig) {\bf 12}, 174 (2003).  
  
  \bibitem{SQC4b}
  W.~Guzman, J.~Socorro, V.~I.~Tkach, and J.~Torres,
   Phys.\ Rev.\ D {\bf 69}, 043506 (2004);
    J.~Socorro and O.~Obregon,
    Rev.\ Mex.\ Fis.\  {\bf 48}, 205 (2002).

\bibitem{Z14}  P. V. Moniz, Phys. Rev. D \textbf{57} R7071 (1998).
  
\bibitem{Swfu1} P.~V.~Moniz,
   {\tt gr-qc/9605034}; 
   A.~Macias, O.~Obregon and J.~Socorro,
   Int.\ J.\ Mod.\ Phys.\ A {\bf 8}, 4291 (1993).
   
  \bibitem{Swfu2}  P. V. Moniz, Int. J. Mod. Phys. D \textbf{6}, 625 (1997); 
  O.~Obregon and C.~Ramirez,
   Phys.\ Rev.\ D {\bf 57}, 1015 (1998).

\bibitem{Swfu3} V.~I.~Tkach, O.~Obregon, and J.~J.~Rosales,
   Class.\ Quant.\ Grav.\  {\bf 14}, 339 (1997);
   V.~I.~Tkach, J.~J.~Rosales and J.~Socorro,
    Mod.\ Phys.\ Lett.\ A {\bf 14}, 1209 (1999);
   O.~Obregon, J.~Socorro, V.~I.~Tkach, and J.~J.~Rosales,
  Class.\ Quant.\ Grav.\  {\bf 16}, 2861 (1999).

\bibitem{KS91}C. Kiefer and T. P. Singh, Phys. Rev. D {\bf 44}, 1067 (1991).
\bibitem{Honnef}C. Kiefer, in {\em Canonical gravity: from classical
 to quantum}, edited by J. Ehlers and H. Friedrich (Springer, Berlin, 1994).
\bibitem{BK} A. O. Barvinsky and C. Kiefer,
Nucl. Phys. B {\bf 526}, 509 (1998).
\bibitem{Lueck} T. L\"uck, The semiclassical approximation
 of supersymmetric quantum gravity. {\em Diplomarbeit}, Universit\"at
zu K\"oln, 2004 (unpublished).
\bibitem{KPS} C. Kiefer, T. Padmanabhan, and T. P. Singh,
Class. Quantum Grav. {\bf 8}, L185 (1991).
\bibitem{GK} D. Giulini and C. Kiefer, Class. Quantum Grav. {\bf 12}, 403
(1995).

\bibitem{Carroll:Friedmann:94}
S.~M. Caroll, D.~Z. Friedmann, M.~E. Ortiz, and D.~N. Page,
Nucl. Phys B {\bf 423}, 661 (1994).
\bibitem{DeWitt:67}
B.~S. DeWitt, Phys. Rev. {\bf 160}, 1113 (1967);
in {\em Relativity}, edited by M. Carmeli, S. I. Fickler, and
L. Witten (Plenum Press, New York, 1970).
\bibitem{Gerlach:69}
U.~H. Gerlach, Phys. Rev. {\bf 177}, 1929 (1969).


\bibitem{AG} P. C. Aichelburg and R. G\"uven, Phys. Rev. Lett. {\bf 51},
 1613 (1983).

 \bibitem{AG1} P. C. Aichelburg and F. Embacher, Phys. Rev. D
  {\bf 34},
 3006 (1986); {\em ibid}. {\bf 37}, 338 (1988).

 \bibitem{AG3} R. Arnowitt and P. Nath, Gen. Rel. Grav. {\bf 7}, 89 (1976).

 \bibitem{AG4} B. DeWitt, {\em SuperManifolds}, second edition
 (Cambridge University Press, Cambridge, 1992).

\bibitem{Pauli} W. Pauli, Helv. Phys. Acta {\bf V}, 179 (1932).

\bibitem{SU} R. U. Sexl and H. K. Urbantke, {\em Relativity, groups,
particles} (Springer, Wien, 2001).
\bibitem{Wulf} M. Wulf, Int. J. Mod. Phys. D {\bf 6}, 107 (1997).
\bibitem{NPZ} H. Nicolai, K. Peeters, and M. Zamaklar, {\tt hep-th/0501114}.
\bibitem{deco} E. Joos, H. D. Zeh, C. Kiefer, D. Giulini, J. Kupsch,
I.-O. Stamatescu, 
{\em Decoherence and the appearance of
 a classical world in quantum theory}, second edition (Springer, Berlin, 2003).
\bibitem{CK88} C. Kiefer, Phys. Rev. D {\bf 38}, 1761 (1988).
\bibitem{LL} L. D. Landau, E. M. Lifshitz, and J. Menzies,
{\em Quantum Mechanics} (Butterworth--Heinemann~1997).

\bibitem{KuHu} S. Sinha and B. L. Hu, Phys. Rev. D {\bf 44}, 1028 (1991); 
K. V. Kucha\v{r} and M. P. Ryan, Jr., Phys. Rev. D {\bf 40}, 3982 (1989). 

\bibitem{SiPa} T. P. Singh and T. Padmanabhan,  Ann. Phys. (N.Y.) 
{\bf 196}, 296 (1989). 

\bibitem{CKpla} D. Giulini and C. Kiefer, Phys. Lett. A 
{\bf 193}, 21 (1994).

 \bibitem{Giulini} D. Giulini, Phys. Rev. D {\bf 51}, 5630 (1995).

\bibitem{vN} P. van Nieuwenhuizen, Phys. Rep. \textbf{68}, 189 (1981).

  \end{thebibliography}
\end{document}